\documentclass[10pt,a4paper]{article}

\usepackage{amsmath,amssymb}
\usepackage{color}

\numberwithin{equation}{section}
\newcommand{\p}{\partial}
\newcommand{\e}{\epsilon}
\newcommand{\al}{\alpha}
\newcommand{\nn}{\nonumber}
\newcommand{\dd}{\mathcal{D}}
\newcommand{\A}{\mathcal{A}}
\newcommand{\cp}{\mathcal{P}}
\newcommand{\tw}{\tilde{w}}

\newtheorem{thm}{Theorem}[section]
\newtheorem{cor}[thm]{Corollary}
\newtheorem{lem}[thm]{Lemma}
\newtheorem{prp}[thm]{Proposition}
\newtheorem{emp}[thm]{Example}
\newtheorem{rmk}[thm]{Remark}
\newtheorem{dfn}[thm]{Definition}
\newtheorem{cnj}[thm]{Conjecture}

\newenvironment{prf}{\noindent {\it Proof} \ }{\hfill $\Box$}
\newenvironment{prfn}[1]{\noindent {\it Proof of #1} \ }{\hfill $\Box$}

\DeclareMathOperator{\res}{res}

\begin{document}

\title{Hamiltonian Structures and Reciprocal Transformations for the $r$-KdV-CH Hierarchy}
\author{Ming Chen\footnote{chen-m02@mails.tsinghua.edu.cn}\ ,
Si-Qi Liu\footnote{liusq@mail.tsinghua.edu.cn}\ , Youjin Zhang\footnote{youjin@mail.tsinghua.edu.cn}\\
{\small Department of Mathematical Sciences, Tsinghua
University}\\
{\small Beijing 100084, P.R. China}}
\date{\today}
\maketitle

\begin{abstract}
The $r$-KdV-CH hierarchy is a generalization of the Korteweg-de Vries and
Camassa-Holm hierarchies parametrized by $r+1$ constants.
In this paper we clarify some properties of its multi-Hamiltonian structures, prove the semisimplicity
of the associated bihamiltonian structures and the formula for their central invariants.
By introducing a class of generalized Hamiltonian structures,
we give in a natural way the transformation formulae
of the Hamiltonian structures of the hierarchy under certain reciprocal transformation,
and prove the formulae at the level of its dispersionless limit. We also
consider relations of the associated bihamiltonian structures to Frobenius manifolds.

\end{abstract}
\tableofcontents
\section{Introduction}
In recent years progress has been made in the study of the problem of
classification of bihamiltonian structures of certain type, the associated
bihamiltonian integrable hierarchies include in particular the well known Korteweg-de Vries (KdV) hierarchy,
the Camassa-Holm (CH) hierarchy,
the Drinfeld-Sokolov hierarchies and so on \cite{CH, CHH, DZ1, DZ2, DLZ1, DLZ2, LZ}. For a given bihamiltonian
structure defined on the formal loop space of an $n$-dimensional manifold $M$,
a complete set of its invariants under the so called Miura-type transformations is obtained in \cite{DLZ1, LZ},
this set consists of
a flat pencil of metrics on the manifold $M$ and $n$ functions of one variable, these functions are called the
central invariants of the bihamiltonian structure. These invariants enable one to have a better understanding
of the bihamiltonian structures and the associated integrable hierarchies. For most of the well known bihamiltonian
integrable
hierarchies including the Drinfeld-Sokolov hierarchies associated to untwisted affine Lie algebras,
the flat pencil of metrics are given by certain Frobenius manifold structures, and the central invariants
are constants. In particular, for the bihamiltonian structures of the Drinfeld-Sokolov hierarchies
associated to the untwisted
affine Lie algebras of A-D-E type, the central invariants are all equal to $\frac1{24}$ if one choose the
invariant bilinear form of the Lie algebra to be the normalized one. This property is one of the most
important characteristics of the integrable hierarchies that arise in 2d topological field theory and Gromov-Witten
invariants\cite{DW, DZ1, DZ2, DLZ2, EYY, konts, Witten2}.

On the other hand, up to our knowledge the only known bihamiltonian integrable
hierarchies with non-constant central invariants
are special cases of the so called $r$-KdV-CH hierarchy (see its
definition given in the next section). This hierarchy is a generalization
of the KdV hierarchy parameterized by an ordered set of $r+1$ constants
$\cp=(a_0,a_1,\cdots,a_r)$. Apart from the KdV hierarchy, it
contains many other important integrable hierarchies as particular
examples, such as the the CH hierarchy, the AKNS hierarchy and the
two-component Camassa-Holm (2-CH) hierarchy. In these cases the corresponding parameters are given by
\begin{align*}
\mbox{KdV}:&\quad r=1,\ \cp=(1,0), \\
\mbox{CH}:&\quad r=1,\ \cp=(0,1), \\
\mbox{AKNS}:&\quad r=2,\ \cp=(1,0,0), \\
\mbox{2-CH}:&\quad r=2,\ \cp=(0,0,1).
\end{align*}
The central invariants
of the bihamiltonian structures of the KdV hierarchy and AKNS hierarchy are constant, while that of the
CH hierarchy and the 2-CH hierarchy are not.

In some cases, bihamiltonian integrable hierarchies with different central invariants are related via certain
type of transformations which change, unlike the Miura-type transformations, also the independent variables. Such
transformations are called reciprocal transformations, they are rather important
in studying properties of solutions of the related integrable hierarchies (see \cite{CLZ, FP1, Fu, LZJ, Pav, RS, XZ} and references therein).
A typical example is given by
the relation between the KdV hierarchy and the
CH hierarchy \cite{Fu}, the associated reciprocal transformation provides an efficient way to obtain exact
solutions of the CH hierarchy by using the known solutions of the KdV hierarchy\cite{Fu, LZJ}.

The main purpose of the present paper is to study, via the example
of the $r$-KdV-CH hierarchy, the transformation rule of Hamiltonian
and bihamiltonian structures under reciprocal transformations. A better understanding
of such transformation rules would be important in particular for the study of properties of the
class of bihamiltonian integrable systems of Camassa-Holm type, and for the study
of a generalized classification scheme for integrable hierarchies under reciprocal transformations.

Recall that the first step towards the generalization of the KdV
hierarchy to the $r$-KdV-CH hierarchy was made by Mart\'inez Alonso
in \cite{MA}, where he presented the $r$-KdV-CH hierarchy with the
parameters $\cp=(1,0,\cdots,0)$ and proved its integrability via the
bihamiltonian structure of the hierarchy and the inverse scattering
method. In \cite{AF2, AF3, AF4} Antonowicz and Fordy studied the
spectral problem associated to the $r$-KdV-CH hierarchy with general
parameters, they obtained $r+1$ Hamiltonian operators associated to
the spectral problem and pointed out that the compatibility of these
Hamiltonian operators can be proved by using Fuchssteiner and Fokas'
method of hereditary symmetry \cite{FF}. However, to our knowledge
the explicit formulation of the $r$-KdV-CH hierarchy and its
multi-Hamiltonian representation, including the proof of the
compatibility of the Hamiltonian structures, were missing in the
literature. So in the present paper we first give, in Section
\ref{sec-2} and Section \ref{sec-3}, the explicit formulation of the
$r$-KdV-CH hierarchy with general parameters $\cp=(a_0, \cdots,
a_r)$ and its $r+1$ Hamiltonian structures including the explicit
formulae of the Hamiltonians, we prove the compatibility of the
$r+1$ Hamiltonian operators by using a direct and simple method due
to Kersten {\em et al} \cite{KKV} and Getzler \cite{Ge}. In Section
\ref{sec-2} we also give a definition of the $\tau$ function of the
$r$-KdV-CH hierarchy, it is a natural generalization of the $\tau$
function of the KdV hierarchy which plays a crucial role in the
study of properties of solutions of the KdV hierarchy and its
applications in different branches of mathematics and physics.

Properties of the bihamiltonian structures of the $r$-KdV-CH
hierarchy were considered in \cite{DLZ1}, where a formula for the
central invariants of these bihamiltonian structures was given
without a proof. In Section \ref{sec-4} we fill the proof for the
formula, and also the proof of the semisimplicity of the
bihamiltonian structures. The dispersionless limit of these
bihamiltonian structures are of hydrodynamic type, such
bihamiltonian structures have close relations to 2d topological
field theory and Frobenius manifolds \cite{Du1, taniguchi,
DZ2, Witten2}. The notion of Frobenius manifolds was invented by Dubrovin as a
coordinate free formulation of the WDVV equations which arises in 2d
topologoical field theory \cite{verdier, Du1}, through this notion relations
of some nonlinear integrable systems with 2d topological field
theory, Gromov-Witten invariants, singularities theory, geometry of
the orbit spaces of finite Coxeter groups and extended affine Weyl
groups, and other fields of mathematical physics are naturally
established. Here bihamiltonian structures of hydrodynamic type
play a prominent role, it was shown that on the loop space of any
Frobenius manifold there is defined such a bihamiltonian structure
\cite{Du1}, and vice versa, under certain restrictions a
Frobenius manifold can be obtained from a bihamiltonian structure
\cite{taniguchi, DZ2}. In Section \ref{sec-4} we will specify those
bihamiltonian structures of the $r$-KdV-CH hierarchy which are
associated to Frobenius manifolds.

Under reciprocal transformations an evolutionary PDE which possesses a
local Hamiltonian structure will in general be transformed to a system
with nonlocal Hamiltonian structures. For a Hamiltonian system
of hydrodynamic type, such transformation properties were studied by
Ferapontov and Pavlov in \cite{FP1} where the nonlocal Hamiltonian structure
was obtained from the expressions of the transformed systems.
In order to understand such transformation rule of the Hamiltonian structures in a more
natural way, we first
generalize in Section \ref{sec-5}, following \cite{reci}, the definition of the
space of multi-vectors on the formal loop space of the manifold $M$, and the
Schouten-Nijenhuis bracket on the space of the generalized multi-vectors.
In this way, we can define a class of generalized Hamiltonian structures which includes in particular the
class of weakly nonlocal Hamiltonian structures of hydrodynamic type associated to conformally flat metrics \cite{fera}.
We proceed to define a class of reciprocal transformations between two spaces of
generalized multi-vectors, and obtain in a natural way the transformation rule
of a local Hamiltonian structure under a class of reciprocal transformations.
Then we consider, by applying the general results, the transformation rule of the Hamiltonian structures of
the $r$-KdV-CH hierarchy under the reciprocal transformations of the
hierarchy.

We give some concluding remarks in the last section.

\section{The $r$-KdV-CH hierarchy}\label{sec-2}

\subsection{Definition of the flows}
Let $M$ be a contractible manifold with local coordinates $w^0,
\cdots, w^{r-1}$, $\varphi$ be a smooth map from circle
$S^1=\mathbb{R}/\mathbb{Z}$ to $M$
\[\varphi:S^1 \to M,\quad x\mapsto(w^0(x),\cdots, w^{r-1}(x)).\]
We denote the derivatives $\p_x w^i(x), \p_x^2 w^i(x), \cdots$ by
$w^{i}_{x}, w^{i}_{xx}, \cdots$, and denote $\p_x^k w^i(x)=w^{i,k}$ in general. Let $\bar{\A}$ be the polynomial
ring
\[\bar{\A}=C^\infty(M)[\e\,w^{i}_{x}, \e^2\,w^{i}_{xx}, \cdots].\]
There is a natural gradation on $\bar{\A}$
\[\deg f(w)=0,\ \deg \left(\e\,w^{i}_{x}\right)=1,\ \deg \left(\e^2\,w^{i}_{xx}\right)=2, \cdots\]
We denote the completion of $\bar{\A}$ w.r.t. this gradation by $\A$.

Given $r\in\mathbb{N}$ and $\cp=(a_0, a_1,\cdots,a_r)\in(\mathbb{R}^{r+1})^{\times}$, we consider the
following system of linear equations:
\begin{align}
\e^2 \phi_{xx}(x,t)&=A(w;\lambda)\phi(x,t),\quad A=\frac{\hat{A}}{{\hat{a}}},\label{lax-x}\\
\phi_t(x,t)&=B(w;\lambda) \phi_x(x,t)-\frac12 B_x(w;\lambda)\phi(x,t), \label{lax-t}
\end{align}
where
\begin{align}
\hat{A}&=w^0+w^1\lambda+\cdots+w^{r-1}\lambda^{r-1}+\lambda^r,\label{z-a-h}\\
\hat{a}&=a_0+a_1\lambda+\cdots+a_{r-1}\lambda^{r-1}+a_r\lambda^r,
\end{align}
and $B(w;\lambda)\in \A[\lambda, \lambda^{-1}]$. The compatibility condition of \eqref{lax-x} and \eqref{lax-t} reads
\begin{equation}
A_t=2\,A\,B_x+A_x\,B-\frac{\e^2}2B_{xxx}. \label{lax-eq}
\end{equation}

For a Laurent series $F=\sum\limits_{i\in\mathbb{Z}}f_i\lambda^i$,
we denote by $(F)_+$ its positive part
$$(F)_+=\sum_{i\ge0}f_i \lambda^i,
$$ and $(F)_-=F-(F)_+$. We assume $w^r=1$ and $w^i=a_i=0$, if $i<0$ or $i>r$, and
define the operators
\[\dd_i=2\,w^i\,\p_x+w^{i,x}-\frac{\e^2}2\,a_i\,\p_x^3.\]
Their generating ``function'' is denoted by
\[\hat{\dd}=\sum_{i=0}^r \dd_i\lambda^i=2\,\hat{A}\,\p_x+\hat{A}_x-\frac{\e^2}2\,\hat{a}\,\p_x^3.\]

\begin{prp}
Suppose $B$ is a polynomial of $\lambda$
\begin{equation}
B(w;\lambda)=\tilde{b}_0 \lambda^n+\tilde{b}_1 \lambda^{n-1}+\cdots+\tilde{b}_{n-1}\lambda+\tilde{b}_n,\ \tilde{b}_i \in \A, \label{bbb}
\end{equation}
then the compatibility condition of \eqref{lax-x} and \eqref{lax-t}
gives an evolutionary PDE of the unknown $w^0, \cdots, w^{r-1}$ if and
only if
\[B(w;\lambda)=(f(\lambda)b)_+,\]
where $f(\lambda)\in\mathbb{R}[\lambda], \deg f=n$, $b$ is a Laurent series
\[b=1+\frac{b_1}{\lambda}+\frac{b_2}{\lambda^2}+\frac{b_3}{\lambda^3}+\cdots,\ b_i \in \A\]
such that
\begin{equation}
\hat{A} b^2-\frac{\e^2}4\hat{a}(2\,b\,b_{xx}-b_x^2)=\lambda^r.
\label{eq-b}
\end{equation}
\end{prp}

\begin{rmk}
According to the definition of $\A$, The equation \eqref{eq-b}
detemines $b$ uniquely. For example, the vanishing of the
coefficient of $\lambda^{r-1}$ implies
\[2b_1+w^{r-1}-\frac{\e^2}2a_r\,b_{1,xx}=0,\]
so we have
\[b_1=\left(1-\frac{\e^2}4a_r\p_x^2\right)^{-1}\left(-\frac{w^{r-1}}2\right)=\sum_{k=0}^\infty\left(\frac{\e^2}4a_r\p_x^2\right)^k\left(-\frac{w^{r-1}}2\right).\]
The other $b_i$'s are similar.
\end{rmk}

\begin{prf}
By comparing the coefficients, the compatibility condition \eqref{lax-eq} is equivalent to the following equations
\begin{align}
\lambda^{r+n}:\quad &0=\dd_r \tilde{b}_0, \label{eq-b0}\\
\lambda^{r+n-1}:\quad &0=\dd_r \tilde{b}_1+\dd_{r-1}\tilde{b}_0, \label{eq-b1}\\
\cdots:\quad &\cdots\cdots\cdots\nn\\
\lambda^{r+1}:\quad &0=\dd_r \tilde{b}_{n-1}+\cdots+\dd_{r-n+1}\tilde{b}_0, \\
\lambda^{r}:\quad &0=\dd_r \tilde{b}_n+\cdots+\dd_{r-n}\tilde{b}_0, \label{eq-bn}\\
\lambda^{r-1}:\quad &(w^{r-1})_t=\dd_{r-1} \tilde{b}_n+\cdots+\dd_{r-n-1}\tilde{b}_0, \label{eq-ww}\\
\lambda^{r-2}:\quad &(w^{r-2})_t=\dd_{r-2} \tilde{b}_n+\cdots+\dd_{r-n-2}\tilde{b}_0, \\
\cdots:\quad &\cdots\cdots\cdots\nn\\
\lambda^{1}:\quad &(w^1)_t=\dd_1 \tilde{b}_n+\dd_0 \tilde{b}_{n-1}, \label{eq-w1}\\
\lambda^{0}:\quad &(w^0)_t=\dd_0 \tilde{b}_n. \label{eq-w0}
\end{align}
They give a system of PDEs of $w^0,\cdots, w^{r-1}$ if and only if
$\tilde{b}_0, \cdots, \tilde{b}_n$ satisfy the equations
\eqref{eq-b0}-\eqref{eq-bn}.

To find out $\tilde{b}_0, \cdots, \tilde{b}_n$ from the equations \eqref{eq-b0}-\eqref{eq-bn}, we only need to consider the following equation
\begin{equation}
\hat{\dd}\tilde{b}=0,
\end{equation}
where
\[\tilde{b}=\tilde{b}_0+\frac{\tilde{b}_1}{\lambda}+\frac{\tilde{b}_2}{\lambda^2}+\frac{\tilde{b}_3}{\lambda^3}+\cdots.\]
It is easy to see
\[0=\tilde{b}\left(\hat{\dd}\tilde{b}\right)=\left(\hat{A}\tilde{b}^2-\frac{\e^2}4\hat{a}\left(2\,\tilde{b}\,\tilde{b}_{xx}-\tilde{b}_x^2\right)\right)_x.\]
Note that the kernel of $\p_x$ on $\A$ is just $\mathbb{R}$, so there exists a Laurent series
\[C(\lambda)=C_0\lambda^r+C_1\lambda^{r-1}+C_2\lambda^{r-2}+C_3\lambda^{r-3}+\cdots,\ C_i\in\mathbb{R},\]
such that
\[\hat{A}\tilde{b}^2-\frac{\e^2}4\hat{a}\left(2\,\tilde{b}\,\tilde{b}_{xx}-\tilde{b}_x^2\right)=C(\lambda).\]
Compare the above equation and the equation \eqref{eq-b}, we obtain
\[\tilde{b}=b\,\sqrt{C_0+\frac{C_1}{\lambda}+\frac{C_2}{\lambda^2}+\frac{C_3}{\lambda^3}+\cdots},\]
so we have
\[B=(\lambda^n \tilde{b})_+=(f(\lambda)b)_+,\]
where
\[f(\lambda)=\left(\lambda^n\sqrt{C_0+\frac{C_1}{\lambda}+\frac{C_2}{\lambda^2}+\frac{C_3}{\lambda^3}+\cdots}\right)_+.\]
The proposition is proved.
\end{prf}

\begin{prp}
Suppose $B$ is a polynomial of $\lambda^{-1}$
\[B=\frac{\tilde{c}_0}{\lambda^m}+\frac{\tilde{c}_1}{\lambda^{m-1}}+\cdots+\frac{\tilde{c}_{m-1}}{\lambda},\ \tilde{c}_i\in\A.\]
The compatibility condition of \eqref{lax-x} and \eqref{lax-t} gives
an evolutionary PDE of unknown $w^0, \cdots, w^{r-1}$ if and only if
\[B=\left(g(\lambda)c\right)_-,\]
where $g(\lambda)\in\mathbb{R}[\lambda^{-1}], \deg g=m$, $c$ is power series
\[c=c_0+c_1\,\lambda+c_2\,\lambda^2+c_3\,\lambda^3+\cdots,\ c_i\in\A\]
such that
\begin{equation}
\hat{A} c^2-\frac{\e^2}4\hat{a}(2\,c\,c_{xx}-c_x^2)=1. \label{eq-c}
\end{equation}
\end{prp}
\begin{prf}
The proof is similar to the above one, we omit it here.
\end{prf}

\begin{prp}
Suppose $B$ is a Laurent polynomial of $\lambda$
\[B=\frac{\tilde{b}_{-m}}{\lambda^m}+\frac{\tilde{b}_{-m+1}}{\lambda^{m+1}}+\cdots+\frac{\tilde{b}_{-1}}{\lambda}+\tilde{b}_0
+\tilde{b}_1\,\lambda+\cdots+\tilde{b}_{n-1}\,\lambda^{n-1}+\tilde{b}_n\lambda^n.\]
The compatibility condition of \eqref{lax-x} and \eqref{lax-t} gives
an evolutionary PDE of unknown $w^0, \cdots, w^{r-1}$ if and only if
\[B=\left(f(\lambda)b\right)_++\left(g(\lambda)c\right)_-,\]
where $f(\lambda)\in\mathbb{R}[\lambda],\ \deg f=n,\ g(\lambda)\in\mathbb{R}[\lambda^{-1}],\ \deg g=m$, and $b$, $c$ are given in the above two propositions.
\end{prp}
\begin{prf}
By comparing the coefficients of $\lambda$, it is easy to see that $\tilde{b}_0, \cdots, \tilde{b}_n$ satisfy the equations \eqref{eq-b0}-\eqref{eq-bn},
and $\tilde{b}_{-m}, \cdots, \tilde{b}_{-1}$ satisfy the same equation for $\tilde{c}_0, \cdots, \tilde{c}_{m-1}$. The proposition is proved.
\end{prf}

\begin{dfn}
For $n\in\mathbb{Z}$, we define
\begin{equation}
B_n=\left\{\begin{array}{ll} (\lambda^n b)_+, & n\ge0, \\ (\lambda^n c)_-, & n<0, \end{array} \right.
\end{equation}
and denote by $\p_{t_n}$ the flow defined by $B_n$. When $n>0$, we call $\p_{t_n}$ the $n$-th positive flow, and when $n<0$, we call
$\p_{t_n}$ the $|n|$-th negative flow.
\end{dfn}

\begin{lem}
For all $n\in\mathbb{Z}$, we have
\begin{align}
b_{t_n}&=B_n\,b_x-b\,B_{n,x}, \label{eq-bt}\\
c_{t_n}&=B_n\,c_x-c\,B_{n,x}. \label{eq-ct}
\end{align}
\end{lem}
\begin{prf}
We only prove the equation \eqref{eq-bt}, the proof of \eqref{eq-ct} is similar.

Let $h=b_{t_n}-B_n\,b_x+b\,B_{n,x}$, we need to show that $h=0$.
The derivative of the equation \eqref{eq-b} w.r.t. $t_n$ implies
\[A_{t_n}\,b^2+\left(2\,A\,b-\frac{\e^2}2\left(b\,\p_x^2-b_x\,\p_x+b_{xx}\right)\right)\,b_{t_n}=0,\]
from \eqref{lax-eq} it follows that
\[b\,B_n\,\left(\hat{\dd}b\right)+\left(2\,A\,b-\frac{\e^2}2\left(b\,\p_x^2-b_x\,\p_x+b_{xx}\right)\right)\,h=0,\]
then by using the equation \eqref{eq-b} again and the fact that $\hat{\dd}b=0$, we obtian
\[\left(1-\frac{\e^2}4\frac{\hat{a}}{\lambda^r}\left(b^2\,\p_x^2-b\,b_x\,\p_x+b_x^2-b\,b_{xx}\right)\right)\,h=0,\]
thus we arrive at $h=0$.
\end{prf}

\begin{thm}
For all $n,m\in\mathbb{Z}$, we have $[\p_{t_n},\p_{t_m}]=0$.
\end{thm}
\begin{prf}
We need to prove that $(A_{t_n})_{t_m}=(A_{t_m})_{t_n}$, but in fact
\[(A_{t_n})_{t_m}-(A_{t_m})_{t_n}=\hat{\dd}\left(B_{n,t_m}-B_{m,t_n}+B_n\,B_{m,x}-B_m\,B_{n,x}\right),\]
so we only need to prove
\begin{equation}
B_{n,t_m}-B_{m,t_n}+B_n\,B_{m,x}-B_m\,B_{n,x}=0. \label{btbt}
\end{equation}

When $n, m\ge0$, we have
\begin{align*}
B_{m,t_n}-B_n\,B_{m,x}&=\left(\lambda^m\,b_{t_n}\right)_+-B_n\,B_{m,x}\\
&=\left(B_n\,\left(\lambda^m\,b\right)_x-\left(\lambda^m\,b\right)\,B_{n,x}\right)_+-B_n\,B_{m,x}\\
&=\left(B_n\,\left(\lambda^m\,b_x\right)_--\left(\lambda^m\,b\right)\,B_{n,x}\right)_+\\
&=\left(\left(\lambda^n\,b\right)\,\left(\lambda^m\,b_x\right)_--\left(\lambda^m\,b\right)\,\left(\lambda^n\,b_x\right)_+\right)_+\\
&=\left(\left(\lambda^n\,b\right)\,\left(\lambda^m\,b_x\right)_-+\left(\lambda^m\,b\right)\,\left(\lambda^n\,b_x\right)_--\lambda^{n+m}\,b\,b_x\right)_+
\end{align*}
Since the above expresion is symmetric w.r.t. $n, m$, the equation \eqref{btbt} is proved.

When $n\ge0, m<0$, note that $B_{n,t_m}$ is a polynomial of $\lambda$, while $B_{m,t_n}$ is a polynomial of $\lambda^{-1}$,
so the equation \eqref{btbt} is equivalent to the following two equations:
\begin{align}
B_{n,t_m}&=\left(B_m\,B_{n,x}-B_n\,B_{m,x}\right)_+, \label{bntm}\\
B_{m,t_n}&=\left(B_n\,B_{m,x}-B_m\,B_{n,x}\right)_-. \label{bmtn}
\end{align}

In fact,
\begin{align*}
B_{n,t_m}&=\left(\lambda^n\,b_{t_m}\right)_+\\
&=\left(B_m\,\left(\lambda^n b_x\right)-\left(\lambda^n b\right)\,B_{m,x}\right)_+\\
&=\left(B_m\,\left(\lambda^n b_x\right)_+-\left(\lambda^n b\right)_+\,B_{m,x}\right)_+\\
&=\left(B_m\,B_{n,x}-B_n\,B_{m,x}\right)_+
\end{align*}
so the equation \eqref{bntm} is proved.

The proofs of the equation \eqref{bmtn} and the case when $n<0, m<0$ are similar, we omit them here.
\end{prf}

The above theorem shows that the flows
$\{\p_{t_n}\}_{n\in\mathbb{Z}}$ form a integrable hierarchy, we call
it the $r$-KdV-CH hierarchy associated to $\cp=(a_0, \cdots, a_r)$.

\subsection{$\tau$ function of the $r$-KdV-CH hierarchy}

\begin{lem}\label{lem-tau}
Let $w^0, w^1, \cdots, w^{r-1}$ be a solution to the $r$-KdV-CH
hierarchy, then the following one-form
\begin{equation}\label{zh-1}
\omega=b_1\,dx+\sum_{n\ge0}b_{n+1}\,dt_n-\sum_{n<0}c_{-n-1}dt_n
\end{equation}
is closed. Here $b_n, c_n$ are defined in \eqref{eq-b}, \eqref{eq-c}.
\end{lem}
\begin{prf}
Since $\p_{t_0}=\p_x$, we can identify $x$ and $t_0$. In the equation \eqref{eq-b},
take the residue at $\lambda=0$, we obtain
\[\frac{\p b_1}{\p t_n}=\frac{\p \omega_n}{\p x}, \mbox{ where }
\omega_n=\left\{\begin{array}{ll} b_{n+1}, & n\ge0, \\ -c_{-n-1}, & n<0. \end{array}\right.\]
To prove that $\omega$ is closed, we only need to show
\begin{equation}
\frac{\p \omega_n}{\p t_m}-\frac{\p \omega_m}{\p t_n}=0,\ \forall n, m \in\mathbb{Z}. \label{otot}
\end{equation}
Note that the left hand side of the above equation is a differential
polynomial of $w^0, w^1, \cdots, w^{r-1}$ with positive degree, and
we have
\[\frac{\p}{\p x}\left(\frac{\p \omega_n}{\p t_m}-\frac{\p \omega_m}{\p t_n}\right)
=\frac{\p}{\p t_m}\left(\frac{\p b_1}{\p t_n}\right)-\frac{\p}{\p t_n}\left(\frac{\p b_1}{\p t_m}\right)=0,\]
so the equation \eqref{otot} holds true. The lemma is proved.
\end{prf}

We assume the domain of $(x,\{t_n\})$ is contractible, then $\omega$ is also exact.

\begin{dfn}
A function $\tau(\cdots, t_{-2}, t_{-1}, x+t_0, t_1, t_2, \cdots)$
is called a $\tau$ function of the $r$-KdV-CH hierarchy associated
to $w^0, w^1, \cdots, w^{r-1}$, if
\[\omega=d\left(\p_x\log \tau\right).\]
\end{dfn}

\begin{emp}
By applying the above general construction to the $r$-KdV-CH
hierarchy with $r=1$ and $\mathcal{P}=(1,0)$, we obtain the KdV
hierarchy. Let $w^0=-u$, the coefficient $b_1$ reads
\[b_1=-\frac{w^0}2=\frac{u}2=\p_x^2\log\tau,\]
so we have $u=2\,\p_x^2\log\tau$, which coincides with the definition of the $\tau$ function of the KdV hierarchy.
\end{emp}

{From} the definition of the $\tau$ function, we see that all the unknown functions
$w^0, w^1,\cdots, w^{r-1}$ can be expressed in terms of $\tau$ and its derivatives w.r.t. the time variables $t_n,
n\in\mathbb{Z}$. So the $r$-KdV-CH hierarchy can be represented as a hierarchy of PDEs for a single
unknown function $\tau$. It is an interesting problem to see whether, as it is true for the KdV hierarchy,
this hierarchy of PDEs for $\tau$ can be written as a hierarchy of bilinear equations in the sense of
Hirota \cite{hirota}.

We remark here that the definition of  the $\tau$ function of  the hierarchy is different from the
one that is given in \cite{DZ2} for a hierarchy of Hamiltonian systems, there the closed 1-form is formed
by the appropriately chosen densities of the Hamiltonians. In the present definition, the coefficients
of the 1-form \eqref{zh-1} are not given by the densities of the Hamiltonians of the $r$-KdV-CH hierarchy
in general.

\section{The multi-Hamiltonian formalism}\label{sec-3}

\subsection{The conserved quantities of the $r$-KdV-CH hierarchy}

\begin{dfn}
Elements of the quotient space
$$
\Lambda=\A/\p_x \A
$$
are called local functionals. Denote
the canonical projection $\A\to\Lambda$  by $f\mapsto\int f\,dx$, then $f$ is called a density of the local functional.
\end{dfn}

\begin{rmk}
For a smooth map $\varphi: S^1\to M$, we can lift it to a map
\[\varphi^{\infty}:S^1 \to J^{\infty}(M),\]
where $J^\infty(M)$ is the infinite jet space of $M$.
Then for any $f\in\A$, we can define a functional on the space $C^\infty(S^1\to M)$ with value in $\mathbb{R}[[\e]]$:
\[\varphi \mapsto \int_{S^1} (\varphi^\infty)^*(f)\,d\mu,\]
where $d\mu$ is a Haar measure on $S^1$. We denote this functional by $\int_{S^1} f\,dx$ for short.

The functional $\int_{S^1} f\,dx$ and the local functional $\int f\,dx$ share many common properties.
For example, if $f_1-f_2=\p_x g$, then by definition, we have
\[\int f_1\,dx=\int f_2\,dx.\]
On the other hand we also have
\[\int_{S^1}f_1\,dx-\int_{S^1}f_2\,dx=\int_{S^1}\p_x g\,dx=\int_{S^1}d\left((\varphi^\infty)^*(g)\right)=\int_{\p S^1}(\varphi^\infty)^*(g)=0.\]
Another important common property is the following formulae of integration by parts
\begin{equation}
\int f\,\p_x g\,dx=-\int g\,\p_x f\,dx,\ \int f\,\p_x^s g\,dx=\int g\,(-\p_x)^s f\,dx. \label{int-parts}
\end{equation}
\end{rmk}

Let us consider a system of evolutionary PDEs of the following form
\begin{equation}
w^i_t=X^i,\quad X^i\in\A. \label{pde-x}
\end{equation}
By using the chain rule, for any $f\in\A$, we have
\[f_t=\sum_{i=0}^{r-1}\sum_{s\ge0}\p_x^s X^i\frac{\p f}{\p w^{i,s}}.\]
So the system of evolutionary PDEs \eqref{pde-x} defines a derivation $\hat{X}$ on $\A$
\[X\mapsto\hat{X}=\sum_{i=0}^r\sum_{s\ge0}\p_x^s X^i\frac{\p}{\p w^{i,s}}.\]

Note that $\hat{X}$ commutes with $\p_x$, so it acts on the quotient space $\Lambda$.
Let $H=\int f\,dx\in\Lambda$, denote the action of $\hat{X}$ on $H$ by $X.H$, then we have
\[X.H=\int \hat{X}(f)\,dx=\int \sum_{i=0}^{r-1} X^i \frac{\delta H}{\delta w^i}\,dx,\]
where
\[\frac{\delta H}{\delta w^i}=\sum_{s\ge0}(-\p_x)^s\frac{\p f}{\p w^{i,s}}\]
is the variational derivative.

\begin{dfn}
The local functional $H=\int f\,dx$ is called a conserved quantity of the system of
evolutionary PDEs \eqref{pde-x} if $X.H=0$, i.e. if
there exists an element $g\in\A$ such that
\[\hat{X}(f)=\p_x(g).\]
The density of a conserved quantity is called a conserved density.
\end{dfn}

\begin{thm}
Define local functionals
\begin{equation}
H_n=\left\{\begin{array}{ll} \int \left(2\res_{\lambda=0}\frac{\lambda^n}{b}\right)\,dx, & n\ge0, \\ & \\
-\int \left(2\res_{\lambda=0}\frac{\lambda^n}{c}\right)\,dx, & n<0, \end{array} \right.
\end{equation}
then $H_n$ are conserved quantities of every flow of the $r$-KdV-CH
hierarchy.
\end{thm}
\begin{prf}
The equation \eqref{eq-bt} and \eqref{eq-ct} show that
\[\left(\frac1{b}\right)_{t_n}=\left(\frac{B_n}{b}\right)_x,\ \left(\frac1{c}\right)_{t_n}=\left(\frac{B_n}{c}\right)_x\]
so the coefficients of $\frac1{b}$ and $\frac1{c}$ should be conserved densities of the flow $\p_{t_n}$.
The theorem is proved.
\end{prf}

The following proposition is useful in the next subsection.

\begin{prp}\label{prp-var}
For $n\in\mathbb{Z}$ and $i=0, 1, \cdots, r-1$, we have
\begin{equation}
\frac{\delta H_n}{\delta w^i}=\left\{\begin{array}{ll}
b_{n+1-(r-i)}, & n\ge0, \\ -c_{-n-1-i}, & n<0, \end{array} \right.
\end{equation}
here we assume $b_0=1$ and $b_k=c_k=0$ when $k<0$.
\end{prp}

\begin{lem}
Let $H=\int f\,dx$ be a local functional, $f_0, \cdots, f_{r-1} \in
\A$, then
\[\frac{\delta H}{\delta w^i}=f_i\]
if and only if for any $X^0, \cdots, X^{r-1}\in\A$ there exists
$\sigma\in\A$ such that
\[\p_t(f)=\sum_{i=0}^{r-1}X^i\,f_i+\p_x(\sigma),\]
where $\p_t$ is given by the equations \eqref{pde-x}.
\end{lem}
\begin{prf}
We only need to show the condition is sufficient. Note that for any $X^i\in\A$ there exists $\sigma\in\A$ such that
\[\p_t(f)=\sum_{i=0}^{r-1}X^i\,\frac{\delta H}{\delta w^i}+\p_x(\sigma),\]
so it is sufficient to prove that for a fixed $Z\in\A$, if for any $X\in\A$ we have $X\,Z\in\p_x\A$, then $Z=0$.

Without loss of generality we can assume that $Z$ is homogeneous.
We take $X=Z$, then by assumption there exists a homogenous element $F$ of $\A$ such that $X\,Z=\p_x F$.
Since $F$ is homogeneous, we can assume
\[F=f(w, w_x, \cdots, w^{(s)}),\]
and there exists an index $i$ such that $\frac{\p F}{\p
w^{i,s}}\ne0$. Thus $\p_x F$ has the form
\[\p_x F=A\,w^{i,s+1}+B,\]
where $A, B\in\A$ are independent of $w^{i,s+1}$. On the other hand,
$Z$ is also a polynomial of $w^{i,s+1}$, assume that the degree of
$Z$ w.r.t. $w^{i,s+1}$ is $m$, then the degree of $Z^2$ is $2m$. So
we obtain
\[2m=1,\]
which is impossible unless $Z=0$.
The lemma is proved.
\end{prf}

\vskip 1em

\begin{prfn}{Proposition \ref{prp-var}}
We give the proof for the case when $n\ge0$, the proof for $n<0$ is similar.

The proposition is equivalent to the following identity
\[\frac{\delta}{\delta w^i}\int \frac2{b}\,dx=\lambda^{i-r}b.\]
According to the above lemma, it is equivalent to prove that for any $X^i\in\A$ there exists $\sigma\in\A$ such that
\[\left(\frac1{b}\right)_t=\frac{\hat{A}_t\,b}{2\,\lambda^r}+\sigma_x,\]
where $\p_t$ is given by \eqref{pde-x} and $\hat A$ is defined in \eqref{z-a-h}.

Let $\eta=\frac1{b}$, the equation \eqref{eq-b} implies
\begin{equation}
\hat{A}\,\eta^2+\frac{\e^2}4\hat{a}\left(2\,\eta\,\eta_{xx}-3\,\eta_x^2\right)=\lambda^r\,\eta^4.
\label{eq-eta}
\end{equation}
By taking the action of $\p_t$, we have
\[\hat{A}_t\eta^2+\left(2\,\hat{A}\,\eta-4\,\lambda^r\,\eta^3+\frac{\e^2}2\hat{a}\left(\eta\,\p_x^2-3\,\eta_x\,\p_x+\eta_{xx}\right)\right)\eta_t=0.\]
Then multiply the left-hand side of the above equation by $\eta$ and eliminate
$\hat{A}\,\eta^2$ by using the equation \eqref{eq-eta}, one obtain
\[\left(1-\frac{\e^2}4\hat{a}\left(\frac1{\eta^2}\p_x^2-\frac{3\eta_x}{\eta^3}\p_x+\frac{3\eta_x^2-\eta\eta_{xx}}{\eta^4}\right)\right)\eta_t
=\frac{\hat{A}_t}{2\lambda^r\eta},\] Note that
\[\frac1{\eta^2}\p_x^2-\frac{3\eta_x}{\eta^3}\p_x+\frac{3\eta_x^2-\eta\eta_{xx}}{\eta^4}=\p_x\frac1{\eta}\p_x\frac1{\eta},\]
so we have
\[\left(\frac1{b}\right)_t=\left(1-\frac{\e^2}4\hat{a}\,\p_xb\,\p_xb\right)^{-1}\left(\frac{\hat{A}_t\,b}{2\lambda^r}\right)=\frac{\hat{A}_t\,b}{2\lambda^r}+\sigma_x,\]
where
\[\sigma=\frac{\e^2}4\hat{a}\,b\,\p_xb\left(\sum_{k=0}^{\infty}\left(\frac{\e^2}4\hat{a}\,\p_xb\,\p_xb\right)^k\right)\left(\frac{\hat{A}_t\,b}{2\lambda^r}\right).\]
The proposition is proved.
\end{prfn}

\subsection{The Hamiltonian structures of the $r$-KdV-CH hierarchy}
In this subsection we recall the definition of Hamiltonian
structures. In order to have a convenient description of the
Hamiltonian structures of the $r$-KdV-CH hierarchy, and to give a
direct proof of their compatibility in the next subsection, we
also recall the description of Hamiltonian structures in terms of
super variables, see \cite{KKV, Ge} for details.

\begin{dfn}
A local $p$-vector is a map $P:\Lambda^p\to\Lambda$ of the following form
\begin{align}
&P(F_1, \cdots, F_p)\nn\\
=&\int\sum_{i_1, \cdots, i_p=0}^{r-1}\sum_{s_1, \cdots,
s_p\ge0}P^{i_1, \cdots, i_p}_{s_1, \cdots, s_p}
\left(\p_x^{s_1}\frac{\delta F_1}{\delta
w^{i_1}}\right)\cdots\left(\p_x^{s_p}\frac{\delta F_p}{\delta
w^{i_p}}\right)\,dx,
\end{align}
where $P^{i_1, \cdots, i_p}_{s_1, \cdots, s_p}\in\A$ such that $P$ is alternative.
We denote the linear space of all $p$-vectors by $\Gamma^p$, and $\Gamma=\bigoplus\limits_{p=0}^\infty \Gamma^p$. In particular, $\Gamma^0=\Lambda$.
\end{dfn}

\begin{rmk}
The space $\Gamma^1$ can be identified with the space of all evolutionary PDEs, since
\[P(F)=\int \sum_{i=0}^{r-1}\sum_{s\ge0}P^i_s \p_x^s\frac{\delta F}{\delta w^i}\,dx=\int \sum_{i=0}^{r-1}\tilde{P}^i \frac{\delta F}{\delta w^i}\,dx,\]
where $\tilde{P}^i=\sum_{s\ge0}\left(-\p_x\right)^s\left(P^i_s\right)\in\A$.
\end{rmk}

Define the maps
\[\theta^s_i:\Lambda\to\A,\quad H\mapsto\p_x^s\frac{\delta H}{\delta w^i},\]
and denote $\hat{\A}=\A\otimes\wedge(V)$, where
\[V=\bigoplus\limits_{i=0}^{r-1}\bigoplus\limits_{s\ge0}\mathbb{R}\,\theta^s_i.\]
Here $\wedge(V)$ is the exterior algebra of $V$. The operators $\frac{\p}{\p\theta_i^s}$ that appear below is defined as follows:
\[\frac{\p}{\p\theta_i^s}\,\theta_{i_1}^{s_1}\wedge\cdots \wedge\theta_{i_m}^{s_m}=\left\{
\begin{array}{l} 0,\quad {\mbox{when}}\ (i,s)\ne (i_k, s_k)\ {\mbox{for}}\ k=1,\cdots, m;\\
(-1)^{k-1} \theta_{i_1}^{s_1}\wedge\cdots \wedge\hat{\theta}_{i_k}^{s_k}\wedge\cdots\wedge\theta_{i_m}^{s_m},\quad {\mbox{when}}\
(i,s)=(i_k,s_k).
\end{array}\right.\]
The derivation $\p_x:\A\to\A$ can be extended to the following one
\[\p_x=\sum_{i=0}^{r-1}\sum_{s\ge0}\left(w^{i,s+1}\frac{\p}{\p w^{i,s}}+\theta^{s+1}_i\frac{\p}{\p \theta^s_i}\right):\hat{\A}\to\hat{\A},\]
we denote $\hat{\Lambda}=\hat{\A}/\p_x\hat{\A}$.
Note that the spaces $\hat{\A}$ and $\hat{\Lambda}$ possess natural gradations, which is induced from the gradation on $\wedge(V)$.

\begin{lem}
There exists an isomorphism of the graded linear spaces \[\Psi:\hat{\Lambda}\to\Gamma.\]
\end{lem}
\begin{prf}
We define the map $\Psi:\hat{\Lambda}^p\to\Gamma^p$ as follow
\begin{equation}
\Psi(P)(F_1, \cdots, F_p)=\int \frac{\p^p P}{\p \theta^{s_1}_{i_1}\cdots\p \theta^{s_p}_{i_p}}\theta^{s_1}_{i_1}(F_1)\cdots\theta^{s_p}_{i_p}(F_p)\,dx,
\label{phip}
\end{equation}
then it is easy to see that $\Psi$ is an isomorphism.
\end{prf}

In \cite{Ge} and \cite{KKV}, Getzler and Kersten {\it et al} define a graded Lie algebra bracket on $\hat{\Lambda}$.
\[[\ ,\ ]:\hat{\Lambda}^p\times\hat{\Lambda}^q\to\hat{\Lambda}^{p+q-1},\ (P, Q)\mapsto[P,Q],\]
where
\begin{equation}
[P,Q]=\int\sum_{i=0}^{r-1}\left(\frac{\delta P}{\delta
\theta_i}\frac{\delta Q}{\delta w^i} +(-1)^p\frac{\delta P}{\delta
w^i}\frac{\delta Q}{\delta \theta_i}\right)\,dx \,.\label{bracket}
\end{equation}
and
\begin{equation}
\frac{\delta P}{\delta\theta_i}=\sum_{s\ge 0} (-\p_x)^s  \frac{\p P}{\p \theta^s_i}.
\end{equation}
We call this bracket the Schouten-Nijenhuis bracket\footnote{
The definitions of $[\ ,\ ]$ in \cite{Ge} and \cite{KKV} are slightly different from the one presented here. They are equivalent to each other
after adjusting sign and gradation.}, it satisfies the following properties:
\begin{align}
&[P,Q]=(-1)^{pq}[Q,P], \label{sb-1}\\
&(-1)^{pk}[[P,Q],R]+(-1)^{qp}[[Q,R],P]+(-1)^{kq}[[R,P],Q]=0, \label{sb-2}
\end{align}
where $P\in\hat{\Lambda}^p,\ Q\in\hat{\Lambda}^q,\ R\in\hat{\Lambda}^k$.

The Schouten-Nijenhuis bracket is also defined on $\Gamma$, since $\hat{\Lambda}\cong\Gamma$.

\begin{dfn}
A local bivector $P\in\Gamma^2$ is called a Poisson bivector if $[P,P]=0$.
Two linearly independent Poisson bivectors $P, Q\in\Gamma^2$ are called to be compatible if $[P,Q]=0$.

A local vector field $X\in\Gamma^1$ is called a Hamiltonian vector field
 w.r.t. a Poisson bivector $P$ if there exists a local functional $H\in\Gamma^0$
such that $X=[P,H]$.
\end{dfn}

Note that the value of a local bivector $P\in\Gamma^2$ on two local functionals can be expressed as follows:
\[P(F,G)=\int\sum_{i,j=0}^{r-1}\frac{\delta F}{\delta w^i}\sum_{s\ge0}
P^{ij}_s\p_x^s\left(\frac{\delta G}{\delta w^j}\right)\,dx,\]
so it gives a matrix differential operator
\[J=(J^{ij})_{i,j=0,\cdots, r-1},\quad J^{ij}=\sum_{s\ge0}P^{ij}_s\p_x^s\,.\]
The fact that $P$ is alternative implies that
\begin{equation}J+J^{\dagger}=0, \label{dagger} \end{equation}
where $J^{\dagger}$ is the adjoint operator of $J$
\[J^{\dagger}=((J^{\dagger})^{ij}),\ (J^{\dagger})^{ij}=\sum_{s\ge0}(-\p_x)^sP^{ji}_s.\]
So the space of local bivectors is isomorphic to the space of matrix differential operators satisfying the condition \eqref{dagger}.
We will identify the matrix differential operators and local bivectors from now on.

Let $P=(P^{ij})=(\sum\limits_{s\ge0}P^{ij}_s\p_x^s)$ be a local bivector, the isomorphism \eqref{phip} implies
\[\Psi^{-1}(P)=\int \left(-\frac12\sum_{i,j=0}^{r-1}\theta_i P^{ij}(\theta_j)\right)\,dx.\]
Let $H\in\Lambda$ be a local functional, and $X=[P,H]$, then we have
\[X=[P,H]=\int\left(\sum_{i,j=0}^{r-1}P^{ij}\left(\frac{\delta H}{\delta w^j}\right)\theta_i\right)\,dx=\int \sum_{i=0}^{r-1}X^i\,\theta_i\,dx.\]

\begin{thm}\label{thm-xph}
Denote by $X_n$ the local vector field corresponding to the
flow $\p_{t_n}$ of the $r$-KdV-CH hierarchy, then there exist $r+1$
Hamiltonian structures $P_m\ (m=0, \cdots, r)$, which are linearly
independent and pairwise compatible, such that
\[X_n=\left\{
\begin{array}{ll}
[P_{r-m}, H_{n+m}], & n\ge0,\\
{[}P_m, H_{n-m}], & n<0,
\end{array}
\right., \ n\in\mathbb{Z},\ m=0, \cdots, r.\]
\end{thm}
\begin{prf}
We will only prove the case when $n\ge0$, the case when $n<0$ can be proved in the same way.

It follows from the equations \eqref{eq-ww}-\eqref{eq-w0} and Proposition \ref{prp-var} that
\begin{equation}
w^i_{t_n}=\sum_{j=0}^{r-1}(P_r)^{ij}\left(\frac{\delta H_n}{\delta
w^j}\right),\ i=0, \cdots, r-1, \label{wt-pn}
\end{equation}
where $P_r$ is a matrix differential operator
\[P_r=\left( \begin{array}{ccccc}
0 & 0 & \cdots & 0 & \dd_0 \\
0 & 0 & \cdots & \dd_0 & \dd_1 \\
\vdots & \vdots & & \vdots & \vdots \\
0 & \dd_0 & \cdots & \dd_{r-3} & \dd_{r-2} \\
\dd_0 & \dd_1 & \cdots & \dd_{r-2} & \dd_{r-1}
\end{array} \right).\]
By using the definition of the Schouten-Nijenhuis bracket we have $X_n=[P_r,
H_n]$.

Note that $b_j$ satisfies the following recurrence relation
\[b_{n-r+1}=-\dd_0^{-1}\left(\dd_1\,b_{n-r+2}+\cdots+\dd_r\,b_{n+1}\right),\]
so we have
\[\left(\begin{array}{c} b_{n-r+1} \\ b_{n-r+2} \\ \vdots \\ b_{n-1} \\ b_n \end{array}\right)=
R_b\left(\begin{array}{c} b_{n-r+2} \\ b_{n-r+3} \\ \vdots \\ b_n \\
b_{n+1} \end{array}\right),\] where the matrix operator $R_b$ reads
\[R_b=\left(\begin{array}{ccccc}
-\dd_0^{-1}\dd_1 & -\dd_0^{-1}\dd_2 & \cdots & -\dd_0^{-1}\dd_{r-1} & -\dd_0^{-1}\dd_r \\
1 & 0 & \cdots & 0 & 0 \\
0 & 1 & \cdots & 0 & 0 \\
\vdots & \vdots & \ddots & \vdots & \vdots \\
0 & 0 & \cdots & 1 & 0
\end{array}\right).\]
Then by using Proposition \ref{prp-var} again, we obtain
\[w^i_{t_n}=\sum_{j=0}^{r-1}(P_{r-1})^{ij}\left(\frac{\delta H_{n+1}}{\delta w^j}\right),\ i=0, \cdots, r-1, \]
where
\[P_{r-1}=P_{r}\,R_b=
\left( \begin{array}{ccccc}
0 & 0 & \cdots & \dd_0 & 0 \\
0 & 0 & \cdots & \dd_1 & 0 \\
\vdots & \vdots & & \vdots & \vdots \\
\dd_0 & \dd_1 & \cdots & \dd_{r-2} & 0 \\
0 & 0 & \cdots & 0 & -\dd_r
\end{array} \right),\]
so we have $X_n=[P_{r-1}, H_{n+1}]$.

Repeating this procedure, we obtain
\[X_n=[P_r, H_n]=[P_{r-1}, H_{n+1}]=\cdots=[P_0, H_{n+r}],\]
where $P_{r-m}=P_r\,R_b^m,\ m=0, 1, \cdots, r$. It is easy to see
all the $P_m$'s are matrix differential operators and linearly
independent. We will prove that they are Hamiltonian operators and
pairwise compatible in the next subsection.
The theorem is proved.
\end{prf}

\subsection{Compatibility of the Hamiltonian structures}

By the definition given in the last subsection, the components of
the matrix differential operators $P_m\ (m=0, 1, \cdots, r)$ have the
following expression
\[(P_m)^{ij}=f^{ij}_m \dd_{i+j+1-m},\ i,j=0, 1, \cdots, r-1,\]
where $f^{ij}_m$ are constants given by
\[f^{ij}_m=\left\{\begin{array}{ll}+1, & i,j<m, \\ -1, & i,j \ge m, \\ 0, & \mbox{otherwise}. \end{array} \right.\]
It is easy to see that $P_m+(P_m)^{\dagger}=0$, so $P_m\in\Gamma^2$. Furthermore, we have the following theorem, which implies that
$P_m$'s are all Hamiltonian operators and pairwise compatible.

\begin{thm}\label{thm-pp}
For any $n, m=0, 1, \cdots, r$, we have $[P_n, P_m]=0$.
\end{thm}

We prove some lemmas first. Let
\[s_k=\left\{\begin{array}{ll} 1, & k\ge0, \\ 0, & k<0. \end{array}\right.\]

\begin{lem} \label{lem-f}$f^{ij}_m=1-s_{i-m}-s_{j-m}$. \end{lem}
The proof of the lemma is obvious.

\begin{lem} \label{lem-ss}
For any integers $i,j,n,m\ge0$, we have
\[s_{i+j+1-n-m}\left(s_{i-n}+s_{i-m}+s_{j-n}+s_{j-m}-2\right)=s_{i-n}s_{j-m}+s_{i-m}s_{j-n}.\]
\end{lem}
\begin{prf}
Consider the generating function
\[F(x,y)=\sum_{i,j\ge0}s_{i+j+1-n-m}\left(s_{i-n}+s_{i-m}+s_{j-n}+s_{j-m}-2\right)\,x^i\,y^j,\]
where $|x|<1, |y|<1$, one can obtain
\begin{align*}
\sum_{i,j\ge0}s_{i+j+1-n-m}s_{i-n}\,x^i\,y^j&=\frac{y^{n+m-1}}{1-y}\,\frac{\left(\frac{x}{y}\right)^n-\left(\frac{x}{y}\right)^{n+m}}{1-\frac{x}{y}}
+\frac{x^{n+m}}{(1-x)(1-y)},\\
\sum_{i,j\ge0}s_{i+j+1-n-m}s_{i-m}\,x^i\,y^j&=\frac{y^{n+m-1}}{1-y}\,\frac{\left(\frac{x}{y}\right)^m-\left(\frac{x}{y}\right)^{n+m}}{1-\frac{x}{y}}
+\frac{x^{n+m}}{(1-x)(1-y)},\\
\sum_{i,j\ge0}s_{i+j+1-n-m}s_{j-n}\,x^i\,y^j&=\frac{y^{n+m-1}}{1-y}\,\frac{1-\left(\frac{x}{y}\right)^m}{1-\frac{x}{y}}
+\frac{x^m\,y^n}{(1-x)(1-y)},\\
\sum_{i,j\ge0}s_{i+j+1-n-m}s_{j-m}\,x^i\,y^j&=\frac{y^{n+m-1}}{1-y}\,\frac{1-\left(\frac{x}{y}\right)^n}{1-\frac{x}{y}}
+\frac{x^n\,y^m}{(1-x)(1-y)},\\
\sum_{i,j\ge0}s_{i+j+1-n-m}\,x^i\,y^j&=\frac{y^{n+m-1}}{1-y}\,\frac{1-\left(\frac{x}{y}\right)^{n+m}}{1-\frac{x}{y}}
+\frac{x^{n+m}}{(1-x)(1-y)},
\end{align*}
so we have
\[F(x,y)=\frac{x^m\,y^n+x^n\,y^m}{(1-x)(1-y)}=\sum_{i,j\ge0}\left(s_{i-n}s_{j-m}+s_{i-m}s_{j-n}\right)\,x^i\,y^j.\]
The lemma is proved.
\end{prf}

\begin{lem}\label{lem-q}
We denote
\begin{equation}
Q^{ijk}_{nm}=f^{i+j+1-m,k}_nf^{ij}_m+f^{i+j+1-n,k}_mf^{ij}_n, \label{qijk}
\end{equation}
then $Q^{ijk}_{nm}$ is symmetric w.r.t. $i,j,k$ for any integers $n,m\ge0$.
\end{lem}
\begin{prf}
By using Lemma \ref{lem-f}, we have
\begin{align*}
Q^{ijk}_{nm}=&2-s_{i-n}-s_{i-m}-s_{j-n}-s_{j-m}-s_{k-n}-s_{k-m}\\
&+s_{k-n}\,s_{i-m}+s_{k-n}\,s_{j-m}+s_{k-m}\,s_{i-n}+s_{k-m}\,s_{j-n}\\
&+s_{i+j+1-n-m}\left(s_{i-n}+s_{i-m}+s_{j-n}+s_{j-m}-2\right).
\end{align*}
Then Lemma \ref{lem-ss} implies
\begin{align*}
Q^{ijk}_{nm}=&2-s_{i-n}-s_{i-m}-s_{j-n}-s_{j-m}-s_{k-n}-s_{k-m}\\
&+s_{k-n}\,s_{i-m}+s_{k-n}\,s_{j-m}+s_{k-m}\,s_{i-n}+s_{k-m}\,s_{j-n}\\
&+s_{i-n}s_{j-m}+s_{i-m}s_{j-n},
\end{align*}
which is symmetric w.r.t. $i,j,k$. The lemma is proved.
\end{prf}

\vskip 1em

\begin{prfn}{Theorem \ref{thm-pp}}
Let us first introduce some elements $\hat{P}_m\in\hat{\Lambda}^2$
\[\hat{P}_m=\frac12\sum_{i,j=0}^{r-1}\theta_i(P_m)^{ij}(\theta_j),\ m=0, 1, \cdots, r-1,\]
then the theorem is equivalent to the identity $[\hat{P}_n, \hat{P}_m]=0$, where $[\ ,\ ]$ is given in \eqref{bracket}.

By the definition of variational derivative, we have
\[\frac{\delta \hat{P}_m}{\delta \theta_\alpha}=\sum_{k=0}^{r-1}f^{\alpha k}_m \dd_{\alpha+k+1-n}(\theta_k),\
\frac{\delta \hat{P}_m}{\delta
w^\alpha}=\sum_{i,j=0}^{r-1}f^{ij}_m\delta^{\alpha}_{i+j+1-m}\theta_i\,\theta_j',\]
where $F'=\p_x(F)$. So the Schouten-Nijenhuis bracket reads
\begin{align*}
[\hat{P}_n, \hat{P}_m]=&\int \sum_{\alpha=0}^{r-1}\left(
\frac{\delta \hat{P}_n}{\delta \theta_\alpha}\frac{\delta \hat{P}_m}{\delta w^\alpha}+
\frac{\delta \hat{P}_n}{\delta w^\alpha}\frac{\delta \hat{P}_m}{\delta \theta_\alpha}\right)\,dx\\
=&\int \sum_{i,j,k=0}^{r-1} Q^{ijk}_{nm}\,\theta_i\,\theta_j'
\dd_{i+j+k+2-n-m}(\theta_k)\,dx,
\end{align*}
where $Q^{ijk}_{nm}$ is given by \eqref{qijk}.

Denote $l=i+j+k+2-n-m$, we can rewrite $\theta_i\,\theta_j'\,\dd_l(\theta_k)$ as
\begin{align*}
\theta_i\,\theta_j'\,\dd_l(\theta_k)=&w^l\,\theta_i\,\theta_j'\,\theta_k'-w^l\,\theta_i\,\theta_j''\theta_k+\frac{\e^2}2\,a_l\,\theta_i\,\theta_j''\theta_k''\\
&-\theta_i'\,\theta_j'\left(w^l\theta_k-\frac{\e^2}2\,a_l\theta_k''\right)+\left(\theta_i\,\theta_j'\left(w^l\theta_k-\frac{\e^2}2\,a_l\theta_k''\right)\right)'.
\end{align*}
Note that the first term of the above expression is anti-symmetric w.r.t $j,k$, so the symmetry of $Q^{ijk}_{nm}$ implies
\[\int \sum_{i,j,k=0}^{r-1} Q^{ijk}_{nm}\,\left(w^l\,\theta_i\,\theta_j'\,\theta_k'\right)\,dx=0.\]
The second, the third and the fourth terms are similar, so we have
\[[\hat{P}_n, \hat{P}_m]=\int \left(\sum_{i,j,k=0}^{r-1} Q^{ijk}_{nm}\,\theta_i\,\theta_j'\left(w^l\theta_k-\frac{\e^2}2\,a_l\theta_k''\right)\right)'\,dx=0.\]
The theorem is proved.
\end{prfn}

\section{Properties of the bihamiltonian structures}\label{sec-4}

\subsection{Semisimplicity of the bihamiltonian structures}\label{sec-41}
\begin{dfn}
We say that a Hamiltonian structure $P$ is of hydrodynamic type if its components have the form
\[P^{ij}=g^{ij}(w)\,\p_x+\Gamma^{ij}_k(w)\,w^k_x,\quad \det(g^{ij}(u))\ne0.\]
A bihamiltonian structure is of hydrodynamic type if both of its Hamiltonian structures have this property.
\end{dfn}

The leading terms of the Hamiltonian structures $P_m\,(m=0, 1, \cdots,
r)$ are all of hydrodynamic type. We denote the corresponding metrics by
$g_m\,(m=0, 1, \cdots, r)$.

\begin{dfn}
We say that a bihamiltonian structure $(P_1, P_2)$ of hydrodynamic type
\[P^{ij}_a=g^{ij}_a(w)\,\p_x+\Gamma^{ij}_{k,a}(w)\,w^k_x+\cdots,\ a=1,2,\]
is  semisimple if the roots of the characteristic equation
\begin{equation}\label{ch-eq}
\det(g_2(w)-\lambda\,g_1(w))=0
\end{equation}
are not constant and pairwise distinct.
\end{dfn}

Ferapontov proved the following theorem:

\begin{thm}[\cite{Fe}]
Let $(P_1, P_2)$ be a semisimple bihamiltonian structure of hydrodynamic type, then the roots of the characteristic equation \eqref{ch-eq}
form a local coordinate system near every point of $M$, and the metrics $g_1, g_2$ are diagonal
in this coordinate system.
\end{thm}

The $r$-KdV-CH hierarchy has $r+1$ compatible Hamiltonian structures, so any
two of them form a bihamiltonian structure. We denote by
$B_{k,l}=(P_k, P_l)$ the bihamitonian structure formed by $P_k$ and
$P_l$, where $k, l=0, 1, \cdots, r$ and $k \ne l$.

Let $\lambda_1, \cdots, \lambda_r$ be the roots of the following polynomial in $\lambda$:
\[P(\lambda)=\lambda^r+w^{r-1}\,\lambda^{r-1}+\cdots+w^1\,\lambda+w^0.\]
We assume that $\lambda_1, \cdots, \lambda_r$ are pairwise distinct,
so $P'(\lambda_i)\ne0, i=1, 2, \cdots, r$.

\begin{thm}\label{bkl}
For any $k, l=0, 1, \cdots, r$ and $k \ne l$, the leading term of
$B_{k,l}$ is a semisimple bihamiltonian structure of hydrodynamic
type.
\end{thm}

By the implicit function theorem, it is easy to see that
\begin{equation}
\frac{\p \lambda_i}{\p w^k}=-\frac{\lambda_i^k}{P'(\lambda_i)}.
\label{dldw}
\end{equation}
Since we have assumed that $\lambda_i\ne\lambda_j$ for any $i, j=1,
\cdots, r$, the Jacobian
\[\frac{\p(\lambda_1, \cdots, \lambda_r)}{\p(w^0, \cdots, w^r)}=(-1)^r\frac{\Delta(\lambda_1, \cdots, \lambda_r)}{\prod_{i=1}^r P'(\lambda_i)}\ne0,\]
where $\Delta(\lambda_1, \cdots, \lambda_r)$ is the Vandermonde
determinant. So $\lambda_1, \cdots, \lambda_r$ form a coordinate
system. We will show below that the metrics $g_m$ are diagonal in
this coordinate system.

\begin{lem}
In the coordinate system $(\lambda_1, \cdots, \lambda_r)$, the
metrics $g_m$ are diagonal, and the diagonal entries read
\[g^{ii}_m(\lambda)=-\frac{2\lambda_i^m}{P'(\lambda_i)}.\]
\end{lem}
\begin{prf}
The components of $g_m$ in the coordinate system $(w^0, \cdots, w^{r-1})$ read
\[g^{kl}_m(w)=2f^{kl}_m w^{k+l+1-m},\]
so we have
\begin{align*}
g^{ij}_m(\lambda)&=\sum_{k,l=0}^{r-1}2\,\frac{\p \lambda_i}{\p w^k} f^{kl}_m w^{k+l+1-m}\frac{\p \lambda_j}{\p w^l}\\
&=\frac2{P'(\lambda_i)P'(\lambda_j)}\sum_{k,l=0}^{r-1}2f^{kl}_m w^{k+l+1-m}\lambda_i^k\lambda_j^l\\
&=\frac2{P'(\lambda_i)P'(\lambda_j)}\left(\sum_{k,l=0}^{m-1}2\,w^{k+l+1-m}\lambda_i^k\lambda_j^l-\sum_{k,l=m}^{r-1}2\,w^{k+l+1-m}\lambda_i^k\lambda_j^l\right)\\
&=\frac2{P'(\lambda_i)P'(\lambda_j)}\left(\sum_{h=0}^{m-1}w^h\lambda_j^{h+m-1}\sum_{k=h}^{m-1}\left(\frac{\lambda_i}{\lambda_j}\right)^k\right.\\
&\qquad\qquad\qquad\qquad\qquad\qquad\left.-\sum_{h=m+1}^Nw_h\lambda_j^{h+m-1}\sum_{k=m}^{h-1}\left(\frac{\lambda_i}
{\lambda_j}\right)^k\right).
\end{align*}

If $i \ne j$, we have
\begin{align*}
g^{ij}_m(\lambda)&=
\frac2{P'(\lambda_i)P'(\lambda_j)}\left(\sum_{h=0}^{m-1}w^h\lambda_j^{h+m}
\frac{\left(\frac{\lambda_i}{\lambda_j}\right)^h-\left(\frac{\lambda_i}{\lambda_j}\right)^m}{\lambda_j-\lambda_i}\right.\\
&\qquad\qquad\qquad\qquad\qquad\left.-\sum_{h=m+1}^Nw_h\lambda_j^{h+m}
\frac{\left(\frac{\lambda_i}{\lambda_j}\right)^m-\left(\frac{\lambda_i}{\lambda_j}\right)^h}{\lambda_j-\lambda_i}\right)\\
&=\frac2{P'(\lambda_i)P'(\lambda_j)(\lambda_j-\lambda_i)}\left(
\lambda_j^m\left(\sum_{h=0}^{m-1}w^h\lambda_i^h+\sum_{h=m+1}^Nw_h\lambda_i^h\right)\right.\\
&\qquad\qquad\qquad\qquad\qquad\left.-\lambda_i^m\left(\sum_{h=0}^{m-1}w^h\lambda_j^h+\sum_{h=m+1}^Nw_h\lambda_h^h\right)\right)\\
&=\frac{2\left(\lambda_j^m\left(-w^m\lambda_i^m\right)
-\lambda_i^m\left(-w^m\lambda_j^m\right)\right)}{P'(\lambda_i)P'(\lambda_j)(\lambda_j-\lambda_i)}\\
&=0.
\end{align*}

If $i=j$, then
\begin{align*}
g^{ii}_m(\lambda)&=\frac2{\left(P'(\lambda_i)\right)^2}\left(\sum_{h=0}^{m-1}w^h\lambda_i^{h+m-1}(m-h)+\sum_{h=m+1}^{r}w^h\lambda_i^{h+m-1}(m-h)\right)\\
&=\frac2{\left(P'(\lambda_i)\right)^2}\left(\lambda_i^{m-1}m\sum_{h=0}^Nw_h\lambda_i^h-\lambda_i^m\sum_{h=1}^r h\,w^h\lambda_i^{h-1}\right)\\
&=-\frac{2\lambda_i^m}{P'(\lambda_i)}.
\end{align*}

The lemma is proved.
\end{prf}

\vskip 1em

\begin{prfn}{Theorem \ref{bkl}}
In the coordinates $(\lambda_1, \cdots, \lambda_r)$, the
characteristic equation \eqref{ch-eq} of the leading term of
$B_{k,l}$ reads
\[\prod_{i=1}^r\frac{2\left(\lambda\,\lambda_i^k-\lambda_i^l\right)}{P'(\lambda_i)}=0,\]
so the roots are
\[u_i=(\lambda_i)^{l-k}, \ i=1, \cdots, r.\]
Since $k \ne l$, $u_i$'s are not constant; and since $\lambda_i$ are distinct, $u_i$ are also distinct. So the leading term of $B_{k,l}$ is semisimple.
\end{prfn}

We denote the diagonal components of $g_k$ in the canonical
coordinates $u_1, \cdots, u_r$ by $f^1, \cdots, f^r$, then it is
easy to obtain
\[g_k^{ii}=f^i=-2(l-k)^2\frac{\lambda_i^{2l-k-2}}{P'(\lambda_i)},\ i=1, \cdots, r,\]
and the diagonal components of $g_l$ in the canonical coordinates read
\[g_l^{ii}=u_i\,f^i=-2(l-k)^2\frac{\lambda_i^{3l-2k-2}}{P'(\lambda_i)},\ i=1, \cdots, r.\]
These formulae are useful in the following parts of this section.

\subsection{The associated Frobenius manifolds}

The notion of Frobenius manifold, which is introduced by Dubrovin in \cite{verdier, Du1}, is the geometric
description of the WDVV associative equation \cite{DVV, Witten1} that arises in 2d topological field theory.
For a given $r$-dimensional Frobenius manifold $M$, let $v^1,\cdots, v^r$ be its flat coordinates near a point
$v_0\in M$ which is so chosen such that $\frac{\p}{\p v^1}$ is the unit vector field. Then the potential
$F=F(v^1,\cdots, v^r)$ as a function of the flat coordinates satisfies the WDVV associative equations
\begin{eqnarray}
&&\frac{\p^3 F}{\p v^1\p v^\alpha\p v^\beta}=\eta_{\al\beta}=\rm{constant},\ \det(\eta_{\al\beta})\ne 0,\nn\\
&&\frac{\p^3 F}{\p v^\alpha\p v^\beta\p v^\xi}\eta^{\xi\sigma}\frac{\p^3 F}{\p v^\sigma\p v^\nu\p v^\mu}=
\frac{\p^3 F}{\p v^\mu\p v^\beta\p v^\xi}\eta^{\xi\sigma}\frac{\p^3 F}{\p v^\sigma\p v^\nu\p v^\al},\nn\\
&&\qquad {\rm{for\ any \ fixed\ indices\ \al,\beta,\nu,\mu}},\ {\rm{here}}\ (\eta^{\al\beta})
=(\eta_{\al\beta})^{-1},\nn\\
&&\p_E F=(3-d) F+{\rm{quadratic \ terms\ in\ }} v.\nn
\end{eqnarray}
Here $E=\sum_{\al=1}^r (d_r v^r+r_\al) \frac{\p}{\p v^\al}$ is the Euler vector field with $d_\al, r_\al$ be
constants, and the constant $d$ is called the charge of the Frobenius manifold.

On the formal loop space of a Frobenius manifold there is a bihamiltonian structure of hydrodynamic type
$(P_{fm}, Q_{fm})$, its components are given by
\begin{eqnarray}
&&P_{fm}^{\al\beta}=\eta^{\al\beta}\p_x,\quad\nn\\
&&Q_{fm}^{\al\beta}=g^{\al\beta}(v)\p_x+\Gamma^{\al\beta}_\gamma\,v^\gamma_x,\nn
\end{eqnarray}
where
$$
g^{\al\beta}(v)=\sum_{\sigma,\gamma,\nu=1}^r (d_\sigma v^\sigma+r_\sigma) \eta^{\al\gamma}\eta^{\beta\nu} \frac{\p^3 F}
{\p v^\sigma\p v^\gamma\p v^\nu}
$$
are the components of the intersection form of the Frobenius manifold, and
$$
\Gamma^{\al\beta}_\gamma=-g^{\al\nu}\, \Gamma^\beta_{\nu\gamma}
$$
is given by the Christoffel symbols of the Levi-Civita connection of the metric $(g_{\al\beta})=(g^{\al\beta})^{-1}$.

If the Frobenius manifold is semisimple, the associated bihamiltonian structure is also semisimple.
Now an interesting question is: Do the leading terms of the bihamiltonian structures
$B_{k,l}$'s of the $r$-KdV-CH hierarchy are associated to some semisimple Frobenius manifolds?

\begin{thm}\label{thm-fro}
The leading term of the bihamiltonian structure $B_{k,l}=(P_k, P_l)$ is given by
the bihamiltonian structure $(P_{fm}, Q_{fm})$ of a Frobenius manifold if and only if $(k,l)=(0,1)$ and $r=1, 2$.
\end{thm}

This theorem is a corollary of the following lemma.
\begin{lem}[\cite{Du1}]
Let $(P_{fm}, Q_{fm})$ be the bihamiltonian structure associated to a semisimple Frobenius manifold.
We denote, in the canonical coordinates $u^1, \cdots, u^r$, the diagonal components of the flat metric $(\eta^{\al\beta})$ by
\[g_1^{ii}=\frac1{\eta_{ii}},\ i=1, \cdots, r.\]
then we have
\begin{equation}
\frac{\p \eta_{ii}}{\p u^j}=\frac{\p \eta_{jj}}{\p u^i},\quad i, j=1, \cdots, r .\label{cond-fro}
\end{equation}
\end{lem}

\begin{prfn}{Theorem \ref{thm-fro}}
For the bihamiltonian structure $B_{k,l}$, the functions $\eta_{ii}$ read
\[\eta_{ii}=-\frac1{2(l-k)^2}\frac{P'(\lambda_i)}{\lambda_i^{2l-k-2}}.\]

For $i \ne j$ we have
\[\frac{\p \eta_{ii}}{\p u^j}=\frac{\p \lambda^j}{\p u^j}\frac{\p \eta_{ii}}{\p \lambda^j}=-\frac{\lambda_j^{1-l+k}\lambda_i^{2-2l+k}}{2(l-k)^3}
\frac{\p P'(\lambda_i)}{\p \lambda_j}.\]
Note that
\[P'(\lambda_i)=\prod_{k\ne i} (\lambda_i-\lambda_k).\]
so we obtain
\[\frac{\p P'(\lambda_i)}{\p \lambda_j}=-\prod_{k\ne i,j} (\lambda_i-\lambda_k),\]
Thus the derivative $\frac{\p \eta_{ii}}{\p u^j}$ reads
\[\frac{\p \eta_{ii}}{\p u^j}=\frac{\lambda_j^{1-l+k}\lambda_i^{2-2l+k}}{2(l-k)^3}\prod_{k\ne i,j} (\lambda_i-\lambda_k).\]

If the leading term of $B_{k,l}$ comes from a Frobenius manifold, the equation \eqref{cond-fro} must hold true, which implies
\begin{equation}
\lambda_i^{1-l}\prod_{k\ne i,j} (\lambda_i-\lambda_k)=\lambda_j^{1-l}\prod_{k\ne i,j} (\lambda_j-\lambda_k). \label{fro-res}
\end{equation}

When $r\ge3$, the equation \eqref{fro-res} can not be true, since the
left hand side depends on $\lambda_i$ while the right hand side
does not. When $r=2$, the equation \eqref{fro-res} becomes
\[\lambda_i^{1-l}=\lambda_j^{1-l},\]
so we have $l=1$.

According to the above analysis, we have only the following four cases to consider:
\[B_{0,1}(r=1),\ B_{1,0}(r=1),\ B_{0,1}(r=2),\ B_{2,1}(r=2).\]
It is easy to verify, by using the explicit form of the potentials of the 1-dimensional and  2-dimensional
Frobenius manifolds given in \cite{Du1}, that only the $B_{0,1}$'s come from Frobenius
manifolds. When $r=1$, the leading term of $B_{0,1}$ corresponds to
the Frobenius manifold with potential
\[F=\frac{(v^1)^3}{12},\quad v^1=w^0.\]
When $r=2$, the leading term of $B_{0,1}$ corresponds to the
Frobenius manifold with potential
\[F=-\frac{(v^1)^2 v^2}4+\frac14 (v^2)^2\log v^2,\]
where $v^1, v^2$ are flat coordinates of the metric $g_0$ given by
\[v^1=w^1,\ v^2=w^0-\frac14{(w^1)^2}.\]

The theorem is proved.
\end{prfn}

\subsection{The central invariants of the bihamiltonian structures}

Let $(\tilde{P}_1, \tilde{P}_2)$ be a bihamiltonian structure of the form
\[\tilde{P}_1=P_1+\e^2 Q_1+\cdots,\ \tilde{P}_2=P_2+\e^2 Q_2+\cdots,\]
and assume that its leading term $(P_1, P_2)$ is a semisimple
bihamiltonian structure of hydrodynamic type. Let $f^i$ be the
diagonal components of the metric $g_1$ in the canonical
coordinates, $Q_a^{ii}$ be the diagonal components of the
coefficient matrix of the third order differential operator $\p_x^3$
of $Q_a$ in the canonical coordinates, where $i=1, \cdots, r$ and
$a=1,2$. Define the functions
\[c_i(u)=\frac{Q_2^{ii}-u_i\,Q_1^{ii}}{3(f^i)^2},\quad i=1,\cdots, r.\]
They are called the central invariants of the bihamiltonian structure $(\tilde{P}_1, \tilde{P}_2)$ .

The notion of central invariants were introduced in \cite{DLZ1, LZ} for semisimple bihamiltonian structures.
Together with the leading terms,
they give a complete set of invariants of a semisimple bihamiltonian structure under the Miura-type
transformations.
\begin{thm}[\cite{DLZ1}]\label{thm-dfm}
Two bihamiltonian structures with a given semisimple bihamiltonian structure of
hydrodynamic type as their leading terms are equivalent,
under the coordinate transformations of the following type (which are called Miura type transformations)
\begin{equation}\label{miura}
w^i\mapsto\tilde{w}^i=w^i+F^i, \mbox{ where } F^i\in \A, \mbox{ and } \left.F^i\right|_{\e=0}=0,
\end{equation}
if and only if their central invariants coincide.
\end{thm}

The central invariants of the bihamiltonian structures $B_{k,l}$ were considered in \cite{DLZ1},
where explicit formulae (see \eqref{cen} below) to compute these invariants were given. We now give
a proof of the formulae.

\begin{thm}
Let $\cp=(a_0, a_1,\cdots,a_r) \in(\mathbb{R}^{r+1})^{\times}$.
Define a polynomial
\[p(\lambda)=a_0+a_1\lambda+\cdots+a_r\lambda^r.\]
Let $B_{k,l}$ be a bihamiltonian structure of the $r$-KdV-CH
hierarchy associated to $\cp$, then the central invariants of
$B_{k,l}$ are given by the following formulae:
\begin{equation}
c_i(u_i)=\frac{p(\lambda_i)}{24(k-l)\lambda_i^{l-1}},\quad i=1,\cdots,
r.\label{cen}
\end{equation}
Here $u_i, \lambda_i$ are defined in Subsection \ref{sec-41}.
\end{thm}
\begin{prf}
We write the Hamiltonian structure $P_m$ as
\[P_m=P_m^{[0]}+\e^2 (Q_m\p_x^3+\cdots)+{\cal O}(\e^4),\]
where the dots represent terms with lower order in $\p_x$. The
components of the tensor $Q_m$ in the coordinates $w^0, \cdots,
w^{r-1}$ read
\[Q_m^{ij}(w)=-\frac12f^{ij}_ma_{i+j+1-m}.\]
So the components of the tensor $Q_m$ in the coordinates $\lambda_1,
\cdots, \lambda_r$ have the expressions
\begin{align*}
Q_m^{ij}(\lambda)&=-\frac12\sum_{k,l=0}^{r-1}\frac{\p\lambda_i}{\p w^k}f^{kl}_ma_{k+l+1-m}\frac{\p\lambda_j}{\p w^l}\\
&=-\frac1{2P'(\lambda_i)P'(\lambda_j)}\sum_{k,l=0}^{r-1}f^{kl}_ma_{k+l+1-m}\lambda_i^k\lambda_j^l.
\end{align*}
We take $i=j$, then
\begin{align*}
Q_m^{ii}(\lambda)&=-\frac1{2\left(P'(\lambda_i)\right)^2}\sum_{k,l=0}^{r-1}f^{kl}_ma_{k+l+1-m}\lambda_i^{k+l}\\
&=-\frac1{2\left(P'(\lambda_i)\right)^2}\left(\sum_{k,l=0}^{m-1}a_{k+l+1-m}\lambda_i^{k+l}-\sum_{k,l=m}^{r-1}a_{k+l+1-m}\lambda_i^{k+l}\right)\\
&=-\frac1{2\left(P'(\lambda_i)\right)^2}\left(\sum_{h=0}^{m-1}(m-h)a_{h}\lambda_i^{h+m-1}-\sum_{h=m+1}^{r}(m-h)a_{h}\lambda_i^{h+m-1}\right)\\
&=-\frac1{2\left(P'(\lambda_i)\right)^2}\sum_{h=0}^{r-1}(m-h)a_{h}\lambda_i^{h+m-1}\\
&=\frac{\lambda_i^mp'(\lambda_i)-m\lambda_i^{m-1}p(\lambda_i)}{2\left(P'(\lambda_i)\right)^2}
\end{align*}
So the diagonal components of $Q_m$ in the canonical coordinates $u_i=\lambda_i^{l-k}$ read
\[Q^{ii}_m(u)=\left(\frac{\p u^i}{\p \lambda^i}\right)^2Q^{ii}_m(\lambda).\]
Then one can obtain the central invariants \eqref{cen} by a simple computation.
\end{prf}

According to Theorem \ref{thm-dfm}, the above theorem shows that the bihamiltonian structures $B_{k,l}$
associated to different $\cp$'s are not equivalent
under Miura type transformations. In particular, there do not exist Miura type transformations that convert the CH hierarchy to the KdV hierarchy.

\section{Reciprocal transformations}\label{sec-5}

\subsection{The reciprocal transformation of the $r$-KdV-CH hierarchy}\label{sec-51}
We now start to consider a class of transformations of the
$r$-KdV-CH hierarchy which, unlike the Miura-type transformations
\eqref{miura}, also involve the independent variable.
\begin{lem}
Let $w^0, w^1, \cdots, w^{r-1}$ be a solution to the $r$-KdV-CH
hierarchy, then the following one-form
\begin{equation}
\alpha=\frac1{c_0}\left(dx+\sum_{n>0}b_n\,dt_n-\sum_{n<0}c_{-n}dt_n\right)
\end{equation}
is closed. Here $b_n, c_n$ are defined in \eqref{eq-b}, \eqref{eq-c}.
\end{lem}
\begin{prf}
In the equation \eqref{eq-ct}, let $\lambda=0$, we obtain
\[\frac{\p}{\p t_n}\left(\frac1{c_0}\right)=\frac{\p}{\p x}\left(\alpha_n\right), \mbox{ where }
\alpha_n=\left\{\begin{array}{ll} \frac{b_n}{c_0}, & n\ge0, \\ -\frac{c_{-n}}{c_0}, & n<0. \end{array}\right.\]
By using the above equation the lemma can be proved in a way that is similar to the one given in
the proof of Lemma \ref{lem-tau}.
\end{prf}

The above lemma shows that for any given solution $w^0, w^1, \cdots,
w^{r-1}$, we can define a set of new coordinates $(y,
\{s_n\}_{n\in\mathbb{Z}})$
\begin{align}
&dy=ds_0=\alpha=\alpha_0\,dx+\sum_{n\ne0}\alpha_n\,dt_{n},\label{reci-y}\\
&ds_n=dt_{-n},\quad n\ne0,\ n\in \mathbb{Z} \label{reci-s}
\end{align}
which is called the reciprocal transformation of the $r$-KdV-CH
hierarchy.

\begin{prp}
Let $w^0, w^1, \cdots, w^{r-1}$ be a solution to the $r$-KdV-CH
hierarchy, and $\phi$ be a solution to the Lax pair of the
$r$-KdV-CH hierarchy
\[\e^2 \phi_{xx}=A\phi,\quad \phi_{t_n}=B_n\phi_x-\frac12 B_{n,x}\phi,\quad n\in\mathbb{Z}.\]
We define
\begin{align*}
&\tilde{\phi}=\frac{\phi}{\sqrt{c_0}},\\
&\tilde{A}=c_0^2\,A-\frac{\e^2}4\left(2\,c_0\,c_{0,xx}-c_{0,x}^2\right), \\
&\tilde{B}_n=\left\{\begin{array}{ll}\frac{B_{-n}}{c_0}-\alpha_{-n}, & n\ne0 \\ 1, &n=0, \end{array}\right.
\end{align*}
then $\tilde{\phi}, \tilde{A}, \tilde{B}$ satisfy
\begin{equation}
\e^2 \tilde{\phi}_{yy}=\tilde{A}\tilde{\phi},\
\tilde{\phi}_{s_n}=\tilde{B}_n\tilde{\phi}_y-\frac12 \tilde{B}_{n,y}\tilde{\phi},\quad n\in\mathbb{Z}.\label{newlax}
\end{equation}
\end{prp}
\begin{prf}
The reciprocal transformation \eqref{reci-y}, \eqref{reci-s} implies
\[\p_y=\p_{s_0}=c_0\,\p_x,\ \p_{s_n}=\p_{t_{-n}}-c_0\,\alpha_{-n}\,\p_x,\quad  n\ne0.\]
The proposition is proved by a straightforward computation.
\end{prf}

The reciprocal transformation is invertible, i.e. we can obtain $x$ as a function of $(y, \{s_n\}_{n\in\mathbb{Z}})$ from the total differential equation
\[dx=c_0\,dy+\sum_{k>0}c_{k}\,ds_k-\sum_{k<0}b_{-k}\,ds_k,\quad d t_n= d s_{-n},\quad n\ne 0,\ n\in\mathbb{Z}. \]

\begin{cor}
Let $w^i(x,t)\,(i=0, \cdots, r-1)$ be a solution to the $r$-KdV-CH
hierarchy associated to $(a_0, a_1, \cdots, a_{r})$. We define
\[v^{r-i}(y,s)=\left[c_0^2\,w^i-\frac{\e^2}4a_i\left(2\,c_0\,c_{0,xx}-c_{0,x}^2\right)\right]_{x\mapsto x(y,s)},\ \tilde{a}_{r-i}=a_i,\]
then the functions $v^0(y,s),\cdots,v^{r-1}(y,s)$ give
a solution to the $r$-KdV-CH hierarchy associated to $\cp'=\{\tilde{a}_0,
\tilde{a}_1, \cdots, \tilde{a}_{r}\}$.
\end{cor}
\begin{prf}
The function $c_0$ satisfies
\begin{equation}
c_0^2\,w^0-\frac{\e^2}4a_0\left(2\,c_0\,c_{0,xx}-c_{0,x}^2\right)=1,
\label{eq-c0}
\end{equation}
so we have $v^r=1$, then the above proposition implies
\[\tilde{A}=\frac{v^0+v^1\,\tilde{\lambda}+\cdots+v^{r-1}\,\tilde{\lambda}^{r-1}+\tilde{\lambda}^r}
{\tilde{a}_0+\tilde{a}_1\,\tilde{\lambda}+\cdots+\tilde{a}_{r-1}\,\tilde{\lambda}^{r-1}+\tilde{a}_r\,\tilde{\lambda}^r},\
\mbox{ where } \tilde{\lambda}=1/\lambda.\]

{From} this $\tilde{A}$ we can construct the generating function $\tilde{b}, \tilde{c}$ by using the equation \eqref{eq-b} and \eqref{eq-c}. By straightforward
computation, one can obtain
\begin{equation}
\tilde{b}=\frac{c}{c_0},\ \tilde{c}=\frac{b}{c_0}, \label{hth}
\end{equation}
and the Laurent polynomials $\tilde{B}_n$ is exactly given by
\begin{equation}
\tilde{B}_n=\left\{\begin{array}{ll} (\tilde{\lambda}^n \tilde{b})_+, & n\ge0, \\ (\tilde{\lambda}^n \tilde{c})_-, & n<0, \end{array} \right.
\end{equation}
so the corollary is proved.
\end{prf}

The above proposition and its corollary show that the reciprocal
transformation converts the $n$-th positive flow of the $r$-KdV-CH
hierarchy associated to $\cp=(a_0, a_1, \cdots, a_{r-1}, a_r)$ to the
$n$-th negative flow of the $r$-KdV-CH hierarchy associated to
$\cp'=(a_r, a_{r-1}, \cdots, a_1, a_0)$.

\subsection{Generalized Hamiltonian structures and their reciprocal transformations}

The reciprocal transformation is quite different from the Miura type
transformations, since it transforms local Hamiltonian structures to
nonlocal ones. We will study this kind of Hamiltonian structures and
their reciprocal transformation in detail in a separate publication
\cite{reci}. In the present subsection, we quote some results from
\cite{reci}, and use them to study the $r$-KdV-CH hierarchy in the
next subsection. Our approach to understand the transformation rule of a Hamiltonian structure
under reciprocal transformations is as follows. We
first generalize the definition of the
space of multi-vectors and the Schouten-Nijenhuis bracket on the formal loop space of the manifold $M$,
and by using the Schouten-Nijenhuis bracket we can define a class of generalized Hamiltonian structures which
includes in particular the
class of weakly nonlocal Hamiltonian structures of hydrodynamic type associated to conformally flat metrics.
We proceed to define a class of reciprocal transformations between two spaces of
generalized multi-vectors, and obtain in a natural way the transformation rule
of a local Hamiltonian structure under a class of reciprocal transformations.

We denote by $\hat{\Gamma}^p=\mathrm{Alt}(\Lambda^p, \Lambda)$ the linear space of alternative multilinear map from $\Lambda^p$ to $\Lambda$,
and let $\hat{\Gamma}^0=\Lambda,\ \hat{\Gamma}^p=0$ when $p<0$.

\begin{thm}[\cite{reci}]
There exists a unique bracket $[\ ,\ ]: \hat{\Gamma}^p\times\hat{\Gamma}^q\to\hat{\Gamma}^{p+q-1}$ satisfying the following conditions:
\begin{align}
&[P,Q]=(-1)^{pq}[Q,P], \label{nsb-1}\\
&(-1)^{pk}[[P,Q],R]+(-1)^{qp}[[Q,R],P]+(-1)^{kq}[[R,P],Q]=0, \label{nsb-2}\\
&[P, F_1](F_2, \cdots, F_p)=P(F_1, \cdots, F_p), \label{nsb-3}
\end{align}
for any $P\in\hat{\Gamma}^p, Q\in\hat{\Gamma}^q, R\in\hat{\Gamma}^k, F_1, F_2, \cdots, F_p\in\Lambda$.
\end{thm}

Note that $\Gamma\subset\hat{\Gamma}$, and the Schouten-Nijenhuis bracket defined over $\Gamma$ satisfies the condition \eqref{nsb-1}-\eqref{nsb-3},
so it must coincides with the bracket defined in the above theorem, so $\Gamma$ is a subalgebra of $\hat{\Gamma}$.

\begin{dfn}
If $P\in\hat{\Gamma}^2$ satisfies $[P,P]=0$, then $P$ is called a generalized Hamiltonian structure.
\end{dfn}

Below, a generalized Hamiltonian structure will also be called a Hamiltonian structure.

\begin{emp}\label{hydro-ghs}
In \cite{FP1}, Ferapontov and Pavlov considered the reciprocal transformation of a system of hydrodynamic type
with a local Hamiltonian
structure, they showed that the transformed system possesses a weakly nonlocal Hamiltonian structure
associated to a conformally flat metric \cite{fera}.

Let $F, G\in \Lambda$ be two local functionals. A weakly nonlocal Hamiltonian structure associated to
a conformally flat metric correspond to an element $P$ of $\hat{\Gamma}^2$
that maps $(F,G)$ to
\[P(F,G)=\int \frac{\delta F}{\delta w^i}P^{ij}\left(\frac{\delta G}{\delta w^j}\right)\,dx\]
where the operator $P$ has the components
\[P^{ij}=g^{ij}(w)\p_x+\Gamma^{ij}_k(w)w^k_x+Z^i_k(w)w^k_x\p_x^{-1}w^j_x+w^i_x\p_x^{-1}Z^j_k(w)w^k_x,\]
and satisfies the condition $[P, P]=0$. Note that $w^i_x\frac{\delta F}{\delta w^i}$
and $w^i_x\frac{\delta G}{\delta w^i}$ can be expressed as total $x$-derivative of elements of $\cal {A}$,
so $P$ is well-defined.
In fact, for a local functional $F=\int f\,dx$, we have
\[\p_x^{-1}\left(w^k_x\frac{\delta F}{\delta w^k}\right)=W(F),\]
where (see \cite{reci})
\[W(F)=f+\sum_{t\ge0}\sum_{s\ge1}(-1)^s\binom{t+s}{s}\p_s^{s-1}\left(w^{i,t+1}\frac{\p f}{\p w^{i,t+s}}\right).\]
So the action of $P$ can be rewritten as
\begin{align*}
P(F,G)=&\int \left(\frac{\delta F}{\delta w^i}\left(g^{ij}(w)\p_x+\Gamma^{ij}_k(w)w^k_x\right)\left(\frac{\delta G}{\delta w^j}\right)\right.\\
&\left.\quad +Z^i_k(w)w^k_x\left(\frac{\delta F}{\delta w^i}W(G)-W(F)\frac{\delta G}{\delta w^i}\right)\right)\,dx,
\end{align*}
which is an element of $\hat{\Gamma}^2$ indeed.
\end{emp}

We begin to define a class of reciprocal transformations on the space of generalized multi-vectors.

The operator $\p_x$ is a derivation of the algebra $\A$. Let $\rho$ be an invertible element of $\A$, then $\p_y=\rho^{-1}\,\p_x$ is also a derivation. We denote
\begin{align*}
&\Lambda_x=\A/\p_x\A,\quad  \hat{\Gamma}^p_x=\mathrm{Alt}(\Lambda^p_x, \Lambda_x),\\
&\Lambda_y=\A/\p_y\A,\quad \hat{\Gamma}^p_y=\mathrm{Alt}(\Lambda^p_y, \Lambda_y).
\end{align*}
Since $\rho$ is invertible, there is an isomorphism from $\Lambda_x$ to $\Lambda_y$
\[\Phi_0: \Lambda_x \to \Lambda_y,\ f+\p_x\A \mapsto \rho^{-1}\,f+\p_y\A,\]
and $\Phi_0$ induces a series of isomorphisms from $\hat{\Gamma}^p_x$ to $\hat{\Gamma}^p_y$
\[\Phi_p: \hat{\Gamma}^p_x \to \hat{\Gamma}^p_y,\ P \mapsto \Phi_p(P),\]
the action of $\Phi_p(P)$ on $Y_1, Y_2, \cdots, Y_p\in\Lambda_y$ is defined by
\[\Phi_p(P)(Y_1, Y_2, \cdots, Y_p)=\Phi_0\left(P\left(\Phi_0^{-1}Y_1, \Phi_0^{-1}Y_2, \cdots, \Phi_0^{-1}Y_p\right)\right).\]
All these $\Phi_{p\ge0}$ are called the reciprocal transformations w.r.t. $\rho$.

\begin{emp}
Let $X\in\Gamma^1$, then it defines an evolutionary PDE
$$w^i_t=X^i.
$$
We assume that $\rho$ is a conserved density of $\p_t$, so we can define the following reciprocal transformation for the
above system:
\[dy=\rho\,dx+\sigma\,dt,\ ds=dt,\]
where $\sigma \in \A$. After this transformation, the original equation becomes
\[w^i_s=\tilde{X}^i=X^i-\sigma w^i_y.\]
Now we take a local functional $\tilde{F}=\int \tilde{f}\,dy\in\Lambda_y$, then the action of $\tilde{X}$ on $\tilde{F}$ reads
\begin{align*}
\tilde{X}(\tilde{F})=&\int \left(X(\tilde{f})-\sigma \tilde{f}_y\right)\,dy=\int \left(X(\tilde{f})+\sigma_y \tilde{f}\right)\,dy \\
=&\int \left(X(\tilde{f})+\frac{\rho_t}{\rho} \tilde{f}\right)\,dy=\int \rho^{-1}\,X(\rho\,\tilde{f})\,dy
=\Phi_1(X)(\tilde{F}).
\end{align*}
So we have $\Phi_1(X)=\tilde{X}$, i.e. our definition of reciprocal transformation coincides with the reciprocal transformation of flows.
\end{emp}

\begin{lem}[\cite{reci}]
For any $P\in\hat{\Gamma}^p_x,\ Q\in\hat{\Gamma}^q_y$, we have
\[[\Phi_p(P), \Phi_q(Q)]=\Phi_{p+q-1}([P,Q]).\]
\end{lem}

\begin{cor}\label{ppp}
If $P\in\hat{\Gamma}^2_x$ is a Hamiltonian structure, then $\Phi_2(P)$ is also a Hamiltonian structure.
\end{cor}
\begin{prf}
The result follows from the equalities
$$[\Phi_2(P), \Phi_2(P)]=\Phi_3([P, P])=0.
$$
The corollary is proved.
\end{prf}

By acting $\p_y$ on $w^i$ repeatedly, we introduce a new system of coordinates
\begin{equation}\label{ncor}
\tw^{i,s}=\p_y^s w^i.
\end{equation}
For example, we have $\tw^{i,1}=\frac1{\rho} w^{i,1}$.
The variational derivative of a local functional $\tilde{F}=\int \tilde{f}\,dy\in\Lambda_y$ w.r.t. the new coordinates is defined as usual
\[\frac{\delta \tilde{F}}{\delta \tw^i}=\sum_{s\ge0}\left(-\p_y\right)^s\frac{\p \tilde{f}}{\p \tw^{i,s}}.\]
Then one can obtain the following important identity.
\begin{lem}\cite{reci}
\begin{equation}
\frac{\delta}{\delta w^i}(\Phi_0^{-1}(\tilde{F}))=
\rho\frac{\delta \tilde{F}}{\delta \tw^i}+\sum_{s\ge0}\left(-\p_x\right)^s\left(\frac{\p \rho}{\p w^{i,s}}W_y(\tilde{F})\right), \label{ident}
\end{equation}
where
\[W_y(\tilde{F})=\p_y^{-1}\left(\tw^i_y\frac{\delta \tilde{F}}{\delta \tw^i}\right).\]
\end{lem}

\begin{emp}
We consider the reciprocal transformation of a Hamiltonian structure of hydrodynamic type
\[P^{ij}=g^{ij}(w)\p_x+\Gamma^{ij}_k(w)\,w^k_x.\]
Let $\rho=\rho(w)$ be a nowhere vanishing function on $M$, then we can define a series of reciprocal transformation $\Phi_{p\ge0}$ w.r.t. $\rho$.
Let $\tilde{F},\tilde{G} \in \Lambda_y$ be two local functionals, then by using the identity \eqref{ident} and the definition of $\Phi_2$, we can obtain
the action of $\Phi_2(P)$ on $(\tilde{F}, \tilde{G})$ as follow
\begin{align}
\Phi_2(P)(\tilde{F},\tilde{G})=&\int \left(\frac{\delta \tilde{F}}{\delta \tilde{w}^i}\left(\tilde{g}^{ij}(\tw)\p_y+\tilde{\Gamma}^{ij}_k(\tw)\tw^k_y\right)
\left(\frac{\delta \tilde{G}}{\delta \tilde{w}^j}\right)\right.\nn \\
&\left.\quad +\tilde{Z}^i_k(\tw)\tw^k_y\left(\frac{\delta \tilde{F}}{\delta \tilde{w}^i}W_y(G)-W_y(F)\frac{\delta \tilde{G}}{\delta \tilde{w}^i}\right)\right)\,dy,
\label{fera-trans}
\end{align}
where
$$
\tilde{g}^{ij}=\rho^2\,g^{ij},\ \tilde{\Gamma}^{ij}_k=-\tilde{g}^{il}\tilde{\Gamma}^j_{lk},
$$
and $\tilde{\Gamma}^j_{lk}$ is the Christoffel symbol of
the Levi-Civita connection of the metric $(\tilde{g}_{ij})=(\tilde{g}^{ij})^{-1}$, the functions $\tilde{Z}^i_k$ are given by
\[\tilde{Z}^i_k=\rho\,\nabla^i\nabla_k\rho-\frac12\left(\nabla_j\rho\right)\left(\nabla^j\rho\right)\delta^i_k,\]
here $\nabla$ is the Levi-Civita connection of $(g_{ij})=(g^{ij})^{-1}$.

The formula \eqref{fera-trans} coincides with Ferapontov and Pavlov's result \cite{fera}.
In our case, $\tilde{F}, \tilde{G}$ can be arbitrary functionals, while in \cite{fera},
Ferapontov  verified this formula for the functionals that are independent of the jet variables.
\end{emp}

\subsection{The reciprocal transformation of the Hamiltonian structures of the $r$-KdV-CH hierarchy}

We have shown that the $r$-KdV-CH hierarchy possesses $r+1$
compatible Hamiltonian structures. In fact, it also has two generalized
Hamiltonian structures. Let $R=P_r \cdot P_{r-1}^{-1}$, and define
$P_{r+k}=R^k\,P_r \ k=1,2$. The components of $P_{r+1}, P_{r+2}$
read
\begin{align*}
P_{r+1}^{ij}&=\dd_{i+j-r}-\dd_i\dd_r^{-1}\dd_j,\\
P_{r+2}^{ij}&=\dd_{i+j-r-1}-\dd_{i-1}\dd_r^{-1}\dd_j-\dd_i\dd_r^{-1}\dd_{j-1}+\dd_i\dd_r^{-1}\dd_{r-1}\dd_r^{-1}\dd_j.
\end{align*}
We need to show that $P_{r+1}, P_{r+2}\in\hat{\Gamma}^2$. Note that
\[\dd_r^{-1}=\left(2\p_x-\frac{\e^2}{2}a_r\p_x^3\right)^{-1}=\frac12\p_x^{-1}+O(\p_x),\]
where $O(\p_x)$ stands for a differential operator, so we have
\[\dd_i\dd_r^{-1}\dd_j=-\frac12w^{i,x}\p_x^{-1}w^{j,x}+O(\p_x),\]
thus $P_{r+1}\in\hat{\Gamma}^2$ (see Example \ref{hydro-ghs}).
Similarly,
\[\dd_i\dd_r^{-1}\dd_{r-1}\dd_r^{-1}\dd_j=w^{i,x}\p_x^{-1}X^j+X^i\p_x^{-1}w^{j,x}+O(\p_x),\]
wherer $X^i$ are differential polynomials, so we also have
$P_{r+2}\in\hat{\Gamma}^2$.

In the subsection \ref{sec-51}, we proved that there exists a
reciprocal transformation converting the $r$-KdV-CH hierarchy
associated to $\cp=(a_0, a_1, \cdots, a_{r-1}, a_r)$ to the
$r$-KdV-CH hierarchy associated to $\cp'=(a_r, a_{r-1}, \cdots, a_1,
a_0)$. We denote the flows, the Hamiltonians and the Hamiltonian
structures of the $r$-KdV-CH hierarchy associated to $\cp'$ by
$\tilde{X}_n$, $\tilde{H}_n$ and $\tilde{P}_m$ respectively. Then by using the
notations introduced in the last subsection, we have
\[\Phi_1(X_n)=\tilde{X}_{-n},\quad n\ne0.\]
By using the definition of $H_n,\ \tilde{H}_n$ and the equation \eqref{hth}, we obtain
\[\Phi_0(H_n)=-\tilde{H}_{-n-2},\quad n\ne-1.\]

Now let us consider the following identity which is proved in Theorem \ref{thm-xph}:
\[X_n=[P_m, H_{n-m}],\quad m=0,\cdots,r,\  n<0.\]
Since the reciprocal transformation preserves the Schouten-Nijenhuis bracket, we have
\[\tilde{X}_{n}=[\Phi_2(P_m), -\tilde{H}_{n+m-2}], \quad n>0.\]
On the other hand, we know by Theorem \ref{thm-xph} that
\[\tilde{X}_{n}=[-\tilde{P}_{r-m+2}, -\tilde{H}_{n+m-2}], \quad m=2,\cdots, r+2,\ n>0.\]
So from the above two expressions of $\tilde{X}_n$ we can formulate the following conjecture:
\begin{cnj} \label{main-conj} For $m=0, 1, \cdots, r+1, r+2$, we have
\[\Phi_2(P_m)=-\tilde{P}_{r+2-m}.\]
\end{cnj}

This conjecture and Corollary \ref{ppp} imply the following corollary:
\begin{cor}
$P_{r+1}$ and $P_{r+2}$ are Hamiltonian structures.
\end{cor}

It seems us that the trouble in proving the conjecture \ref{main-conj}, if it holds true, lies in the fact that the
reciprocal transformation is defined in terms of  $c_0$ which does not have an
explicit expression. In this paper, we only prove the coincidence of
the leading terms of $\Phi_2(P_m)$ and $-\tilde{P}_{r+2-m}$.

The
leading terms $P_m^{[0]}$ of $P_m\,(m=0, 1, \cdots, r+1, r+2)$ read
\begin{align*}
(P_m^{[0]})^{ij}&=f^{ij}_m\left(2w^{i+j+1-m}\p_x+w^{i+j+1-m}_x\right),\ (m=0, 1, \cdots, r-1, r),\\
(P_{r+1}^{[0]})^{ij}&=2\left(w^{i+j-r}-w^iw^j\right)\p_x+\left(w^{i+j-r}-w^iw^j\right)_x+\frac12w^i_x\p_x^{-1}w^j_x,\\
(P_{r+2}^{[0]})^{ij}&=2\,U^{ij}\p_x+U^{ij}_x-V^i\p_x^{-1}w^j_x-w^i_x\p_x^{-1}V^j,
\end{align*}
where
\begin{align*}
U^{ij}=&w^{i+j-r-1}-w^{i-1}w^j-w^iw^{j-1}+w^{r-1}w^iw^j,\\
V^i=&\frac12w^{i-1}_x+\frac12w^iw^{r-1}_x+\frac14w^{r-1}w^i_x.
\end{align*}

\begin{thm} For $m=0, 1, \cdots, r+1, r+2$, we have
\[\Phi_2(P_m^{[0]})=-\tilde{P}_{r+2-m}^{[0]}.\]
\end{thm}
\begin{prf}
We give here the proof of the theorem for the case when $m=0$, the proof for the other cases are similar.
In the dispersionless case, the function $\rho$ is given by
\[\rho=\frac1{c_0}=\sqrt{w^0},\]
so the formula \eqref{fera-trans} implies
\begin{equation}
-\Phi_2(P_0^{[0]})^{ij}(w)=2\,w^0\,w^{i+j-1}\p_y+\left(w^0\,w^{i+j-1}\right)_y+Z^{i}\p_y^{-1}w^j_y+w^i_y\p_y^{-1}Z^j,
\label{uu}
\end{equation}
where
\[Z^i=\frac{w^1\,w^i_y}{4w^0}+\frac{w^{i+1}\,w^0_y}{2w^0}-\frac{w^{i+1}_y}2.\]
Here since we are considering the dispersionless limit, we do not need to use the notation $\tw^{i,s}=\p_y^s w^i$
to distinguish it from $w^{i,s}$.

The formula \eqref{uu} gives the components of $\Phi_2(P_0^{[0]})$
under the coordinates $w^i$. To prove the theorem, we need to
transform the components to the coordinates
$v^{r-i}=\frac{w^i}{w^0}$, so
\[-\Phi_2(P_0^{[0]})^{ij}(v)=-\frac{\p v^i}{\p w^k}\Phi_2(P_0^{[0]})^{kl}(w)\frac{\p v^j}{\p w^l},\]
where
\[\frac{\p v^i}{\p w^k}=\frac1{w^0}\delta_{k,r-i}-\frac{w^{r-i}}{(w^0)^2}\delta_{k,0}.\]
Then the theorem is proved by a straightforward computation.
\end{prf}

\section{Conclusion}\label{sec-6}
In this paper, we associate to any set $\cp=(a_0, a_1, \cdots, a_r)\in(\mathbb{R}^{r+1})^{\times}$ an integrable hierarchy called the $r$-KdV-CH hierarchy.
We first clarify its multi-Hamiltonian structures, propose a definition of its $\tau$ function, prove the semisimplicity
of the associated bihamiltonian structures and the formulae for their central invariants, and specify the bihamiltonian
structures that are related to Frobenius manifolds. We then introduce the space of generalized multi-vectors
on the formal loop space of the manifold $M$, define the Schouten-Nijenhuis bracket on this space and a class of reciprocal transformations on this space.
In this way, we define a class of generalized Hamiltonian structures, including the weakly nonlocal Hamiltonian structures of hydrodynamic type associated
to conformally flat metrics on $M$. By using the notion of generalized Hamiltonian structures, we give in a natural way the transformation formulae
of the Hamiltonian structures of the $r$-KdV-CH hierarchy under certain reciprocal transformation of the hierarchy,
and prove the formulae at the level of its dispersionless limit.

The problem of how to find solutions to the $r$-KdV-CH hierarchy is still open for a general parameter set $\cp$.
When $\cp=(1,0)$ and $\cp=(1,0,0)$, the associated hierarchies are the well known KdV and AKNS hierarchy
respectively, the methods of solving these hierarchies of evolutionary PDEs are well studied.
Note that these two hierarchies are also the only particular cases among the general $r$-KdV-CH hierarchies
that possess bihamiltonian structures of topological type\footnote{We say a
semisimple bihamiltonian is of topological type, if its leading term
is associated to a Frobenius manifold, and its central
invariants are equal constants.}.
So we may imagine that the rich properties of these two integrable hierarchies are due to
their close relations to topological field theory.
This fact may also give an explanation on why some well known solving methods suitable for these two
hierarchies can not be applied to the $r$-KdV-CH hierarchy associated to
a general parameter set.

When $\cp=(0,1)$ and $\cp=(0,0,1)$, we obtain the CH and $2$-CH hierarchy respectively.
We can  find solutions of these two hierarchies via certain
reciprocal transformations, since they are transformed to the KdV and the AKNS hierarchy after
such transformations\cite{CLZ, Fu,LZJ}. However, we do not
know how to find
exact solutions for the general $r$-KdV-CH hierarchy.

We can regard the $r$-KdV-CH hierarchy as an \textit{energy dependent generalization} of the KdV hierarchy.
It is naturally to ask:
{\em{Whether one can perform similar generalizations to other integrable hierarchies that possess Lax pair
representations}}?
For example, the $x$-part of the Lax pair for the Gelfand-Dickey hierarchy reads
\[\left(\p_x^{n+1}+u_{n-1}\p_x^{n-1}+\cdots+u_1\p_x+\left(u_0-\lambda\right)\right)\phi=0,\]
if we replace it by
\begin{equation}
\left(\left(\sum_{i=0}^ra_i\,\lambda^i\right)\p_x^{n+1}+\sum_{k=0}^{n-1}\left(\sum_{i=0}^r w_{i,k}\lambda^i\right)\p_x^k\right)\phi=0, \label{biglax}
\end{equation}
where $a_i$'s are constants, can we obtain some interesting integrable systems?
Moreover, note that the Gelfand-Dickey hierarchy is the Drinfeld-Sokolov hierarchy associated to the
affine Lie algebra $A_n^{(1)}$ and the fixed vertex $c_0$ of the Dynkin diagram,
so can we formulate a similar question for the Drinfeld-Sokolov hierarchy associated to general $(\mathfrak{g}, c_k)$?

We have the following two examples of such generalization.

Recently the Degasperis-Procesi equation draws much attentions \cite{DHH},
it has the form
\[m_t+3\,m\,u_x+m_x\,u=0,\ m=u-u_{xx}.\]
Its Lax pair reads
\begin{align*}
&\lambda\,\phi_{xxx}-\lambda\,\phi_x+m\,\phi=0, \\
&\phi_t+\lambda\,\phi_{xx}+u\,\phi_x-\left(u_x+\frac{2\lambda}3\right)\phi=0,
\end{align*}
whose $x$-part is exactly of the form \eqref{biglax}. It is well-known that, after a reciprocal transformation, the Degasperis-Procesi equation is transformed to
the first negative flow of the Kaup-Kupershmidt hierarchy, which is the Drinfeld-Sokolov hierarchy
associated to $(A_2^{(2)}, c_1)$.
So we can regard the Degasperis-Procesi equation as a generalization of the Drinfeld-Sokolov hierarchy
associated to the affine Lie algebra $A_2^{(2)}$ and its vertex $c_1$.

The second example is the Sawada-Kotera hierarchy \cite{SK}, which is the Drinfeld-Sokolov hierarchy
associated to $(A_2^{(2)}, c_0)$. There is also a reciprocal transformation
that transforms the first negative flow of this hierarchy to the Novikov equation\cite{HW}
\[m_t+3\,m\,u\,u_x+m_x\,u^2=0,\quad m=u-u_{xx}.\]
The Lax pair of the Novikov equation given in \cite{HW} is of matrix form, which is equivalent to a scalar form similar to \eqref{biglax}.
So the Novikov equation can also be viewed as a generalization of the Drinfeld-Sokolov hierarchy
associated to the affine Lie algebra $A_2^{(2)}$ and the vertex $c_0$ of its associated Dynkin diagram.

These examples support a positive answer to the above questions, we hope to return to them in subsequent
publications.

\vskip 0.4truecm \noindent{\bf Acknowledgments.}
The work is partially supported by the
National Basic Research Program of China (973 Program)  No.2007CB814800
and the NSFC No.10631050.

\end{document}